\documentclass[a4paper,11pt]{article}
\pdfoutput=1 % if your are submitting a pdflatex (i.e. if you have
\usepackage{jcappub} % for details on the use of the package, please
\usepackage[T1]{fontenc} % if needed
\usepackage{amsmath}
\usepackage{amsmath}

\usepackage{graphicx}
\usepackage{color}
\usepackage{natbib}
\usepackage{longtable,lscape}
\usepackage{txfonts}
\usepackage{amsfonts}
\usepackage{amssymb}
\usepackage{newlfont}
\usepackage{textcomp}
\usepackage{times}
\usepackage{units}
\usepackage{mathbbol}
\usepackage[switch]{lineno}
\usepackage{epstopdf}
\usepackage{ifthen}
\def\ba{\begin{eqnarray}}
\def\ea{\end{eqnarray}}
\def\be{\begin{equation}}
\def\ee{\end{equation}}
\newcommand{\comm}[1]{}
\renewcommand{\comm}[1]{ \textcolor{magenta}{#1}}

\newcommand{\dthree}[1]{d^3 \! #1}

\title{\boldmath Constraints on primordial magnetic fields from magnetically-induced perturbations: current status and future perspectives with LiteBIRD and future ground based experiments.}

\author[a,b,1]{D. Paoletti,\note{Corresponding author.}}
\author[a,b]{F. Finelli,}

\affiliation[a]{Osservatorio di Astrofisica e Scienza dello Spazio di Bologna/Istituto Nazionale di Astrofisica, via Gobetti 101, I-40129 Bologna, Italy}
\affiliation[b]{Istituto Nazionale Di Fisica Nucleare, Sezione di Bologna,Viale Berti Pichat, 6/2, I-40127 Bologna, Italy}

\emailAdd{daniela.paoletti@inaf.it}
\emailAdd{fabio.finelli@inaf.it}

\abstract{We present the constraints on the amplitude of magnetic fields generated prior to the recombination using CMB temperature and polarization anisotropy data from Planck 2018 release, alone and in combination with those from BICEP2/Keck array and the South Pole Telescope. We model the fields with a generic parametrization and we make no assumptions on their origin in order to provide general constraints on their characteristics. The analysis updates the former corresponding Planck 2015 results both on data and numerical implementation. We then perform forecasts for the next generation of CMB experiments 
such as LiteBIRD satellite alone and in combination with future ground based experiments. }

\begin{document}
\maketitle
\flushbottom
\section{Introduction}
The origin and evolution of cosmic magnetism is one of the most intriguing topics in modern cosmology. In recent years there has been an impressive development from both data and theoretical studies, and future experiments at different wavelenghts, from the radio to the gamma rays, hold the promise to disclose the many still open issues in the field and provide a complete scenario of cosmic magnetism in the next decade.  The possibility to generate magnetic fields in the early Universe represents a very promising scenario, from the first ideas in the far past \cite{Thorne:1967zz}, up to the present days \cite{Enqvist:1998fw,Olinto:1997sj,Durrer:2006pc,Subramanian:2009fu,Widrow:2011hs,Ryu:2011hu,Caprini:2011cw,Durrer:2013pga,Subramanian:2015lua,Kahniashvili:2018mzl,Subramanian:2018xlb,Subramanian:2019jyd} and it has twofold implications for cosmology. 

On one side these primordial fields can provide the initial seeds for amplification by structure formation and self-induced dynamos. Therefore they can represent the progenitors (or co-progenitors if we consider also astrophysical mechanisms) of the magnetic fields observed in large scale structures like galaxy clusters \cite{Bonafede:2010wg,Govoni:2004as,Carilli:2001hj} and galaxies, the latter up to very high redshift \cite{Kronberg:1993vk,Bernet:2008qp}. A primordial origin might also be a natural explanation for those large scale magnetic fields diffuse in voids, that can be responsible of the anomalous GeV photon deficit in high energy blazar observations \cite{Neronov:1900zz,Taylor:2011bn,Vovk:2011aa,Tavecchio:2010ja,Tavecchio:2010mk,Takahashi:2013lba}. In such a context, future gamma-ray experiments as the Cherenkov Telescope Array, that will also provide time delay and halo measurements of the blazar emission, will strongly improve the lower limits on the amplitude of magnetic fields in voids and at the same time provide hints on a possible helical components \cite{Dermer:2010mm,Long:2015bda,Finke:2013tyq}. On the other end of the observational frequency range, future high sensitivity radio observations will improve significantly the measurements of the magnetic fields in the cosmic structures \cite{Bull:2018lat,OSullivan:2018adp,Loi:2017fge,Govoni:2015ska,Bonafede:2015pka} and will also have the capability to probe the presence of weak fields in the filaments of the large scale structure \cite{Donnert:2018lbe,Giovannini:2015hfa}. 

On the other side, primordial magnetic fields would provide an unconventional glimpse of the physics of the early Universe. 
%Indeed, the generation of primordial magnetic fields would imply significant deviations from the standard models 
%describing the early Universe. 
Several mechanisms could be responsible for the generation of the manetic seeds and we describe the main ones without pretending to be exhaustive. 
%The inflationary mechanisms \cite{Turner:1987bw,Ratra:1991bn,Giovannini:2000dj,Tornkvist:2000js,Bamba:2003av,Ashoorioon:2004rs,Demozzi:2009fu,
%Kanno:2009ei,Caldwell:2011ra,Jain:2012jy,Fujita:2015iga} provide fields with a large coherence lenght but to produce non-negligible amplitudes 
%it requires the breaking of conformal invariance that modifies the dilution of the magnetic fields with the expansion allowing for larger 
%amplitudes at the end of the inflation.
The breaking of conformal invariance during inflation \cite{Turner:1987bw,Ratra:1991bn,Giovannini:2000dj,Tornkvist:2000js,Bamba:2003av,Ashoorioon:2004rs,Demozzi:2009fu,
Kanno:2009ei,Caldwell:2011ra,Jain:2012jy,Fujita:2015iga} can provide seeds with a large coherence lenght but in general a small amplitude. 
Although the amplitude of the seeds can be further amplified during the preheating process after inflation \cite{Finelli:2000sh,Bassett:2000aw,Byrnes:2011aa}, 
issues as back reaction or strong coupling problem need to be addressed during inflation \cite{Kanno:2009ei,Demozzi:2009fu,Bartolo:2013msa}. 
Note that large scale magnetic seeds are also predicted in models which are alternative to inflation \cite{Gasperini:1995dh}.  
%but to produce non-negligible amplitudes
%it requires the breaking of conformal invariance that modifies the dilution of the magnetic fields with the expansion allowing for larger
%amplitudes at the end of the inflation.
%The post-inflationary/causal generation mechanism usually implies 
%the existence of 
PMF could be an imprint of a first order phase transition \cite{Quashnock:1988vs,Vachaspati:1991nm,Baym:1995fk,
Sigl:1996dm,Hindmarsh:1997tj,Grasso:1997nx,Ahonen:1997wh,Boyanovsky:2005ut,Caprini:2009yp,Tevzadze:2012kk,Zhang:2019vsb,Ellis:2019tjf} 
that through bubble nucleation provides the turbulent enviroment necessary to generate and amplify the fields.
%but the standard model is described by simple cross overs and not first order transitions both at EW and QCD scales. 
As an alternative scenario, fields generated through second order perturbations 
via the Harrison mechanism and at recombination \cite{Fidler:2015kkt,Fenu:2010kh,Matarrese:2004kq} are generally too weak to generate 
relevant magnetic seeds but are anyway interesting \cite{Hutschenreuter:2018vkr} for their evolution up to present days in the local Universe. 
Another possibility is that primordial magnetic fields could have been generated at later times during reionization by Biermann battery effect \cite{Gnedin:2000ax}. 
%The inflationary mechanisms \cite{Turner:1987bw,Ratra:1991bn,Giovannini:2000dj,Tornkvist:2000js,Bamba:2003av,Ashoorioon:2004rs,Demozzi:2009fu,
%Kanno:2009ei,Caldwell:2011ra,Jain:2012jy,Fujita:2015iga} provide fields with a large coherence lenght but to produce non-negligible amplitudes 
%it requires the breaking of conformal invariance that modifies the dilution of the magnetic fields with the expansion allowing for larger 
%amplitudes at the end of the inflation. 
%Although many mechanisms like coupling with pseudoscalars and vectors etc., 
%can break the conformal invariance, the change in the fields dilution induces 
%some issues like back reaction or strong coupling problem 
%\cite{Kanno:2009ei,Demozzi:2009fu,Bartolo:2013msa}. To overcome these issues is necessary 
%to introduce modifications to the simplest model of inflation resulting in a much more complex scenario. 
%Alternative models taking place in different times have been proposed from a generation during reionization by Biermann battery effect \cite{Gnedin:2000ax} 
%to generation due to extra dimensions \cite{Kunze:2005ef}.

%Through the understanding of  PMFs characteristics thanks to 
Current and future data will have the possibility to shed light on PMFs characteristics and on their connection to the physics of the early Universe.
%on the physics of the early Universe 
%because the 
PMFs characteristics are indeed related to their generation mechanism, e.g. causal fields are bonded to positive spectral indices, greater 
or equal two \cite{Durrer:2003ja}, in the power law formalism, whereas for inflationary fields the spectral index is connected to the coupling and 
can also assume negative values (with a natural lower bound $n_{B}>-3$ to avoid the introduction of an infrared cut off in the convolution integrals). 
Helicity of PMFs is also important, since is useful 
to sustain an efficient dynamo amplification and avoid saturation. It is possible to generate 
fields with an helical contribution by several mechanisms \cite{Field:1998hi,Vachaspati:2001nb,Sigl:2002kt,Sharma:2018kgs} and numerical 
simulations show how the fields rapidly reach the maximally helical confinguration independentely from the initial conditions 
\citep{Christensson:2000sp,Christensson:2002xu,Saveliev:2013uva,Kahniashvili:2016bkp,Brandenburg:2016odr}. 
Last, but not least, helicity can leave imprints onto the Cosmic Microwave Background 
\cite{Pogosian:2002dq,Caprini:2003vc,Kunze:2011bp,Ballardini:2014jta,Kahniashvili:2014dfa,Ade:2015cva}.

PMFs affect the entire history of the Universe, several effects may be used to constrain their amplitude. PMFs affect the predictions of Big Bang Nucleosynthesis \cite{Kernan:1995bz,Cheng:1996yi,Giovannini:1997gp,Kahniashvili:2010wm,Luo:2018nth}, of the ionization and thermal history \cite{Jedamzik:1999bm,Sethi:2004pe,Tashiro:2006uv,Sethi:2008eq,Schleicher:2008hc,Sethi:2009dd,Schleicher:2011jj,Kunze:2013uja,Pandey:2014vga,Kunze:2014eka,Chluba:2015lpa,Ade:2015cva,Trivedi:2018ejz,Paoletti:2018uic}, of the growth of structure \cite{Tashiro:2009hx,Kahniashvili:2012dy,Fedeli:2012rr,Pandey:2012ss,Kunze:2013hy,Camera:2013fva,Minoda:2018hiu,Safarzadeh:2019kyq} and also other astrophysical observations \cite{Schleicher:2009zb,Hackstein:2019abb,Saga:2018ont}. All these effects provide constraints on the amplitude of the fields spanning from microgauss to the sub-nanoGauss with different level of assumptions on the fields configuration and the theoretical modelling.

Magnetic fields generated prior to the recombination leave distinctive imprints on the Cosmic Microwave Background (hereafter CMB), that is therefore an excellent observable to constrain their origin and characteristics. Being the CMB mainly described by linear physics an advantage with respect to other probes is the possibility to perform accurate theoretical predictions without the use of non-linear simulations. The CMB imprints are several and can affect different aspects of statistics of CMB anisotropies, either breaking isotropy by an homogeneous magnetic field \cite{Barrow:1997mj,Grasso:1996kk,Adamek:2011pr} with the generation of also harmonic space correlations \cite{Durrer:1998ya,Naselsky:2004gm}, or by a stochastic background of PMFs introducing non-Gaussianities with a non-vanishing bispectrum \cite{Brown:2006wv,Seshadri:2009sy,Caprini:2009vk,Shiraishi:2010yk,Trivedi:2010gi,Shiraishi:2011xvp,Shiraishi:2011fi,Shiraishi:2011dh,Shiraishi:2012sn,Shiraishi:2012rm,Shiraishi:2013wua,Ade:2015cva} and trispectrum \cite{Trivedi:2011vt,Trivedi:2013wqa}. 
In the following we will consider only the model of a stochastic background which is compatible with the Lemaitre Friedmann Robertson Walker cosmology.
%As the most likely type of field configuration generated by the abovementioned mechanisms  we consider from now on the model of a stochastic background of PMFs which is supported in a Lemaitre Friedmann Robertson Walker background.
            
The presence of magnetic fields affects the CMB anisotropies angular power spectrum in both temperature and polarization. The different imprints come from different mechanisms where PMFs play a dominant role. The dissipation of PMFs after recombination significantly affects T and E mode polarization through the modification of the ionization history \cite{Kunze:2014eka,Chluba:2015lpa,Ade:2015cva,Paoletti:2018uic}. This effect is very powerful in constraining the PMFs amplitude but is based on several assumptions which are necessary to perform the calculations, future numerical simulations on the interested scales will shed light on the robustness of these assumptions. The presence of PMFs rotates the CMB polarization plane generating B-modes from E-modes and vice-versa in the Faraday rotation effect. Differently from the others PMFs effect on CMB anisotropies the Faraday rotation is coloured, it has a observational frequency dependence as the frequency to minus four that makes it mostly relevant for the lowest frequencies \cite{Kosowsky:1996yc,Kosowsky:2004zh,Campanelli:2004pm,Pogosian:2013dya}. Current high sensitivity observations are mainly oriented to the high frequencies with high sensitivity low frequency observations limited to large angular scales where the Faraday signal is lower. For this reason current Faraday constraints are not competitive with other effects being in the rage of hundreds of nGauss \cite{Kahniashvili:2008hx,Pogosian:2011qv,Ade:2015cva}, but Faraday rotation represents a very good target for future ground based observatories like the QUIJOTE and STRIP experiments \cite{RubinoMartin:2008dg,Incardona:2018fdv} with a good potential also for small scale experiments on higher frequencies thanks to the very high sensitivities \cite{Pogosian:2018vfr} but all must first account for decontamination from the Galactic Faraday rotation which currently sits at the 10 nG level \cite{Oppermann:2011td}. 

The other main effect of CMB angular power spectra is given by the gravitational effect of PMFs which contrary to the other strong effects relies only on minimal assumptions. PMFs gravitate at the level of cosmological perturbations and generate magnetically-induced perturbations exciting scalar, vector and tensor modes, with different initial conditions: compensated \cite{Giovannini:2004aw,Finelli:2008xh,Paoletti:2008ck}, passive \cite{Lewis:2004ef,Shaw:2009nf,Shaw:2010ea} and inflationary\cite{Bonvin:2011dt,Bonvin:2013tba}. 
In this paper we will focus on this effect focusing on the compensated and passive initial conditions, being the inflationary strictly dependent on the choice of generation mechanism, which has been widely explored in the literature \cite{Subramanian:1998fn,Mack:2001gc,Subramanian:2002nh,Subramanian:2003sh,Lewis:2004ef,Giovannini:2004aw,Giovannini:2005jw,Giovannini:2007qn,Finelli:2008xh,Paoletti:2008ck,Shaw:2009nf,Shaw:2010ea,Paoletti:2010rx,Bonvin:2010nr,Paoletti:2012bb,Pogosian:2012jd,Ade:2015cva,Ade:2015cao,Zucca:2016iur,Sutton:2017jgr,Pogosian:2018vfr,Renzi:2018dbq} updating previous results \cite{Ade:2015cva} to the new Planck 2018 data \cite{Aghanim:2019ame} and providing the forecasts for the CMB experiments in the next decade. We will describe the improvement of the treatment we have developed with respect to previous works \cite{Paoletti:2010rx,Paoletti:2012bb,Ade:2015cva} and its impact on the results.

The paper is organized as follows. In section 2 we describe the fields model, in section 3 we provide the theoretical predictions of the CMB angular power spectra. In section 4 we describe the unpdates in methodology and in section 5 we present the results of Planck 2018. Section 6 is dedicated to the forecasts for 
LiteBIRD, in section 7 we present its combination with future ground based experiments. In section 8 we draw our conclusions. 
  
\section{Primordial Magnetic Fields}
We assume a stochastic background of PMFs described by the two point correlation function:
\footnote{We use the Fourier convention:$ Y ({\vec k}{,}\tau) =  \int \dthree{x} \, \mathrm{e}^{i {\vec k} \cdot {\vec x}} Y ({\vec x}{,}\tau) $, and its inverse
$Y ({\vec x}{,}\tau) = \int \frac{\dthree{k}}{(2 \pi)^3} \, \mathrm{e}^{-i {\vec k} \cdot {\vec x}} Y ({\vec k}{,}\tau)$.}
\be
\Big\langle B_a(\vec{k}) \, B_{\mathrm B}^*(\vec{q}) \Big\rangle = \frac{(2\pi)^3 }{2} \, \delta^{(3)}(\vec{k}-\vec{q}) \left[ (\delta_{ab}-\hat{k}_a\hat{k}_{\mathrm B}) \, P_{\mathrm B}(k) + i \, \epsilon_{abc} \, \hat{k}_c \, P_{\mathrm H}(k) \right]\,.
\label{2point}
\ee
We defer the analysis of the helical component for a following work, and therefore we retain only the non-helical term (i.e. $P_{\mathrm H}=0$).
We assume the following scale dependence for the fields power spectrum: $P_{\mathrm B}(k)= A_{\mathrm B} k^{n_{\mathrm B}}$ and, as usual, we adopt the smoothed amplitude of the fields on a Mpc scale:
\ba
B^2_{\mathrm {1\, Mpc}}= \int_0^{\infty} \frac{d{k \, k^2}}{2 \pi^2} \, \mathrm{e}^{-k^2 (1\, {\mathrm {Mpc}})^2} P_{\mathrm B} (k) = \frac{A_{\mathrm B}}{4 \pi^2 \lambda^{n_{\mathrm B}+3}} \, \Gamma \left( \frac{n_{\mathrm B}+3}{2} \right)\,.
\label{gaussian}
\ea 
Magnetic perturbations survive Silk damping thanks to the overdamped mode induced by Alfven velocity:  the fields are damped at smaller scales,  with the exact scale depending on the Alfven velocity and therefore on the amplitude and configuration of the fields. We model this  damping with a sharp cut off in the PMFs power spectrum at the scale $k_{\mathrm D}$ defined as \cite{Subramanian:1997gi,Mack:2001gc}:
 \be
k_\mathrm{D}=(5.5 \times 10^4)^{\frac{1}{n_{\mathrm B}+5}} \left(\frac{B_{1\, {\mathrm {Mpc}}}}{\mathrm{nG}}\right)^{-\frac{2}{n_{\mathrm B}+5}} 
\left(2\pi \right)^{\frac{n_{\mathrm B}+3}{n_{\mathrm B}+5}} h^{\frac{1}{n_{\mathrm B}+5}} \left(\frac{\Omega_{\mathrm b} h^2}{0.022}\right)^{\frac{1}{n_{\mathrm B}+5}} \mathrm{Mpc}^{-1}\,,
\label{kd_def}
\ee
where $h$ is the reduced Hubble constant, $H_0 = 100\,h \, \mathrm{km}\,\mathrm{s^{-1}}\,\mathrm{Mpc^{-1}}$, and $\Omega_{\mathrm b}$ is the baryon density parameter.
The magnetic energy momentum tensor is given by:
\ba
{\mathcal T}^0_0=-\rho_{\mathrm B}=-\frac{B^2({\vec x})}{8\pi a^4(\tau)}\,,\\
{\mathcal T}_i^0=0\,,\\
{\mathcal T}_j^i=\frac{1}{4\pi a^4(\tau)}
\left(\frac{B^2({\vec x})}{2}\,\delta_j^i-B_j({\vec x})\,B^i({\vec x})\right)\,.
\ea
It contributes as an additional source term in the perturbed Einstein equations $\delta G_{\mu\nu}=8\pi G \,(\delta T_{\mu\nu}^{\mathrm Fluid}+\mathcal T_{\mu\nu})$.
Decomposing into scalar, vector and tensor we obtain the different source terms for magnetically induced perturbations. 
\ba
&&\left|\rho_{\mathrm B}(k)\right|^2=\frac{1}{1024\,\pi^5}\int_\Omega \dthree{p} \, P_{\mathrm B}(p) \, P_{\mathrm B}(|{\vec k}-{\vec p}|) \, (1+\mu^2)\,,
\label{density}\\
&&\left|L^{(\mathrm{S})} (k)\right|^2 =\frac{1}{1024\,\pi^5}\int_\Omega \dthree{p} \, P_{\mathrm B}(p) P_{\mathrm B}(|{\vec{k}}-{\vec{p}}|) \, \left[1 + \mu^2 + 4 \gamma \beta(\gamma \beta - \mu)\right] \,,
\label{spectrum_LF}\\
&&\big|\sigma(k)\big|^2 =\frac{1}{1024\,\pi^5} \int_\Omega d^3pP_{\mathrm B}(p)P_{\mathrm B}(|\vec{k}-\vec{p}|)\left[\frac{4}{9}(4+\mu^2-3\gamma^2-3\beta^2+9\gamma^2\beta^2-6\gamma\beta\mu)\right]\nonumber\\
&&\left|\Pi^{(\mathrm{V})}(k)\right|^2=\frac{1}{512\,\pi^5}\int_\Omega \dthree{p} \, P_{\mathrm B}(p)\, P_{\mathrm B}(|{\vec k}-{\vec p}|)\left[(1+\beta^2)(1-\gamma^2) + \gamma\beta(\mu-\gamma\beta)\right] \,,
\label{vector}\\
&&\left|\Pi^{(\mathrm{T})}(k)\right|^2=\frac{1}{512\,\pi^5} \int_\Omega \dthree{p} \, P_{\mathrm B}(p) \, P_{\mathrm B}(|{\vec k}-{\vec p}|)(1+2\gamma^2+\gamma^2\beta^2) \,,
\label{tensor}
\ea
where $\mu = \hat {\vec p} \cdot ({\vec k} -{\vec p})/|{\vec k} -{\vec p}|$,
$\gamma= \hat {\vec k} \cdot \hat {\vec p}$,
$\beta= \hat {\vec k} \cdot ({\vec k} -{\vec p})/|{\vec k} -{\vec p}|$, and $\Omega$ denotes the volume with $p<k_\mathrm{D}$.
The scalar mode has three different source terms: the magnetic energy density, the anisotropic stress and the Lorentz force induced on baryons. Only two of three are independent imposing the energy conservation of the fields. 
This relation allows a freedom of basis for the initial conditions, namely the choice of which two quantities are computed directly from the field configuration for the analysis. We choose the energy density and Lorentz force as independent quantities for the compensated mode, deriving the anisotropic stress. For the passive mode instead we consider the anisotropic stress as the independent quantity, it being the only quantity required.
We assume full anticorrelation between the Lorentz force and the energy density as predicted to lowest order \cite{Paoletti:2010rx}.  
The exact solutions to the convolutions Eq. \ref{tensor} has been derived in \cite{Finelli:2008xh,Paoletti:2008ck} and we refer to them.

The vector mode is source by the anisotropic stress and the Lorentz force whereas the tensor perturbations are sourced solely by the tensor projection of the anisotropic pressure.
 
\section{Magnetically-induced perturbations}
As described in previous sections PMFs source all types of perturbations, with different initial conditions. Scalar and tensor modes are characterized by both compensated and passive initial conditions. The former is the standard particular solution to the inhomogeneous Einstein-Boltzmann equations sourced by PMFs. The term ``compensated'' refers to the compensation between the energy density of the relativistic component of the fluid and the magnetic energy density, and to the compensation between the neutrino anisotropic stress and the magnetic fields's one \cite{Lewis:2004ef,Finelli:2008xh,Paoletti:2008ck}. The latter, the passive modes, derive from the homogeneous solutions before neutrino decoupling when PMFs provide the only anisotropic stress and source both scalar and tensor pre-neutrino-decoupling modes. Both modes are suppressed after the neutrino decoupling thanks to the neutrino compensation described above but a trace of their existence remains as relic inflationary-like modes whose primordial power spectrum is provided by the matching of initial conditions at the decoupling Eq.\ref{tensor} \cite{Lewis:2004ef,Shaw:2009nf,Shaw:2010ea} (and is dependent on PMFs anisotropic stress Fourier spectrum). 
Contrary to scalar and tensors, the vector modes excite only compensated initial conditions.\footnote{A third kind of initial condition has been proposed, called the inflationary mode \cite{Bonvin:2011dt,Bonvin:2013tba}, this is a mode characterizing only the inflationary produced fields and although with a spectral shape similar to the passive case the characteristic of the mode strongly depend on the coupling choice. Since we want to maintain the maximum generality concerning the possible origin of the fields we are not considering this type of initial condition in the current analysis and leave it for a future study.}   

\begin{figure*}[!htb]
\includegraphics[width=0.5\columnwidth]{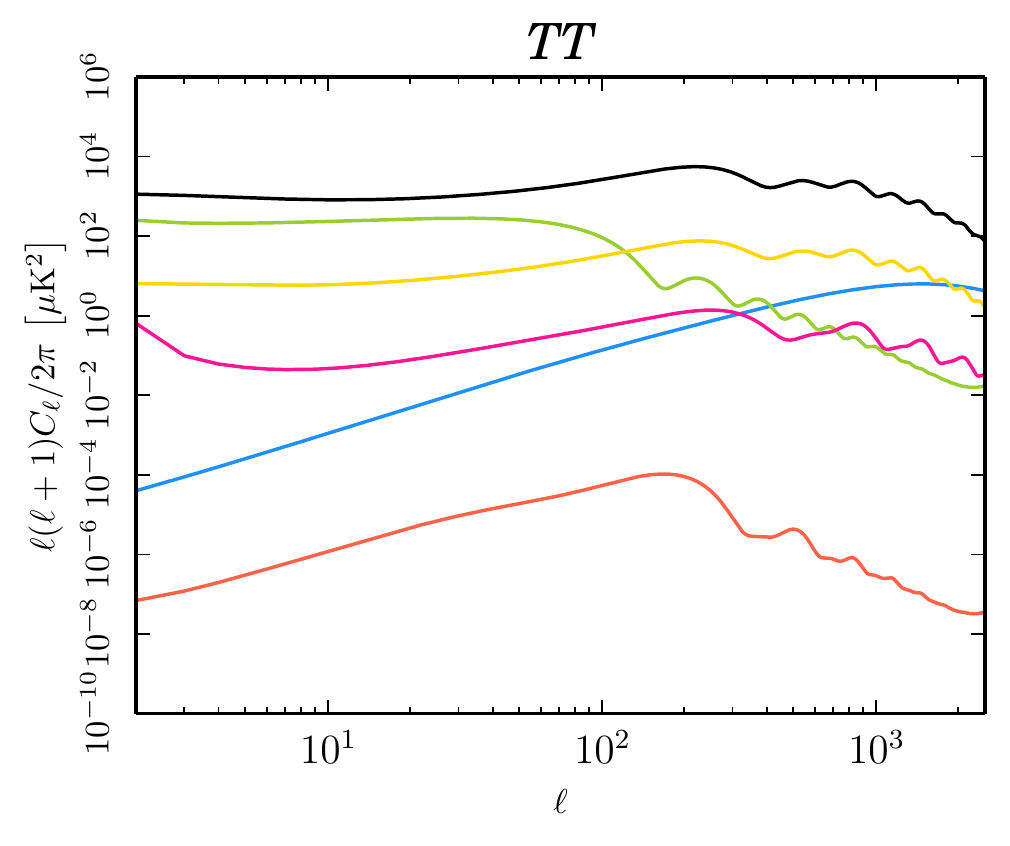}
\includegraphics[width=0.5\columnwidth]{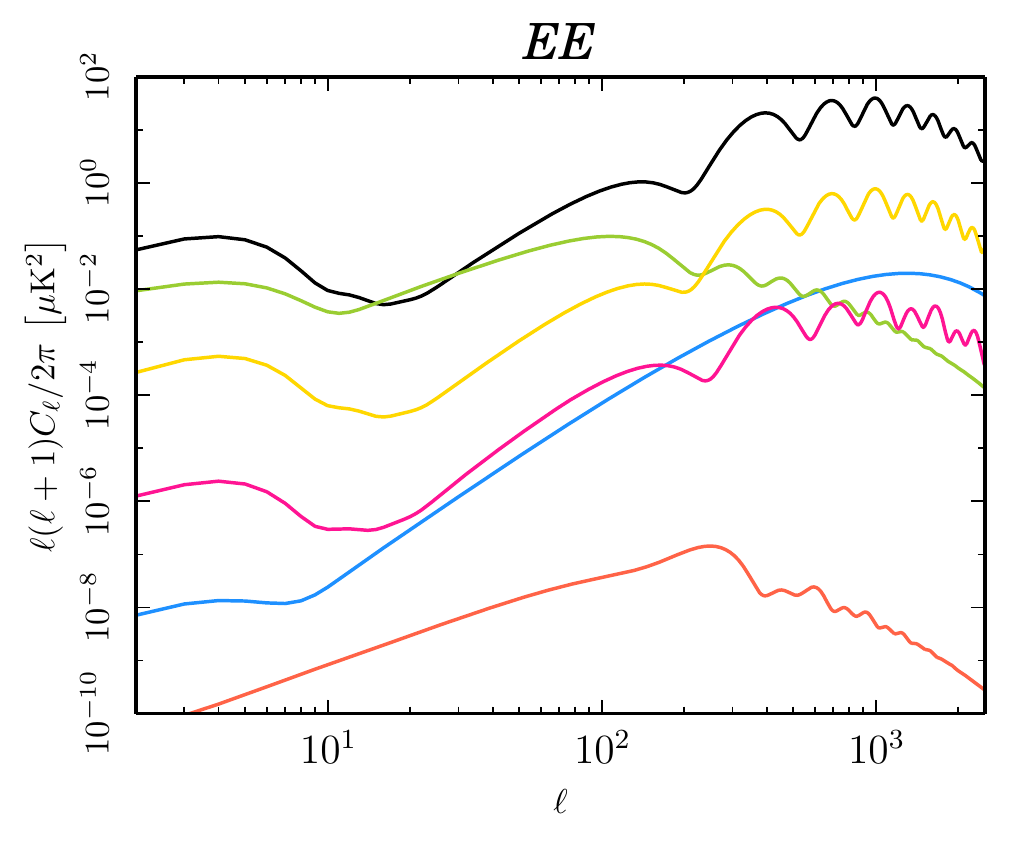}\\
\includegraphics[width=0.5\columnwidth]{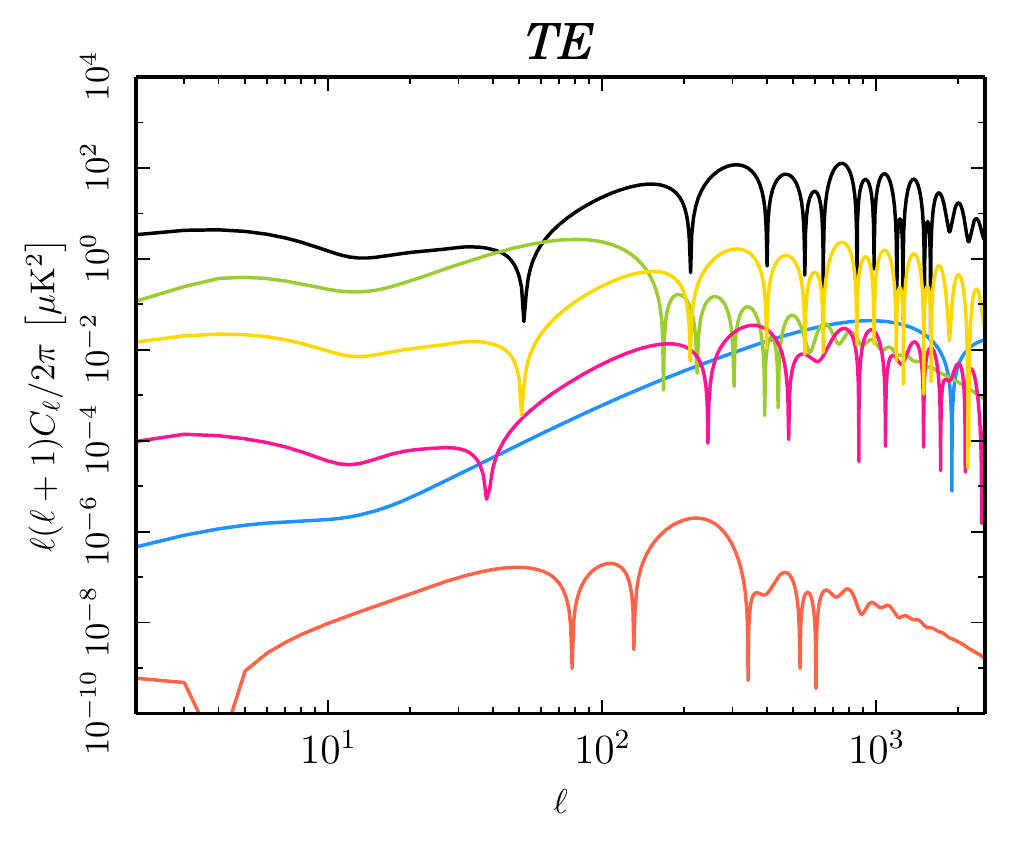}
\includegraphics[width=0.5\columnwidth]{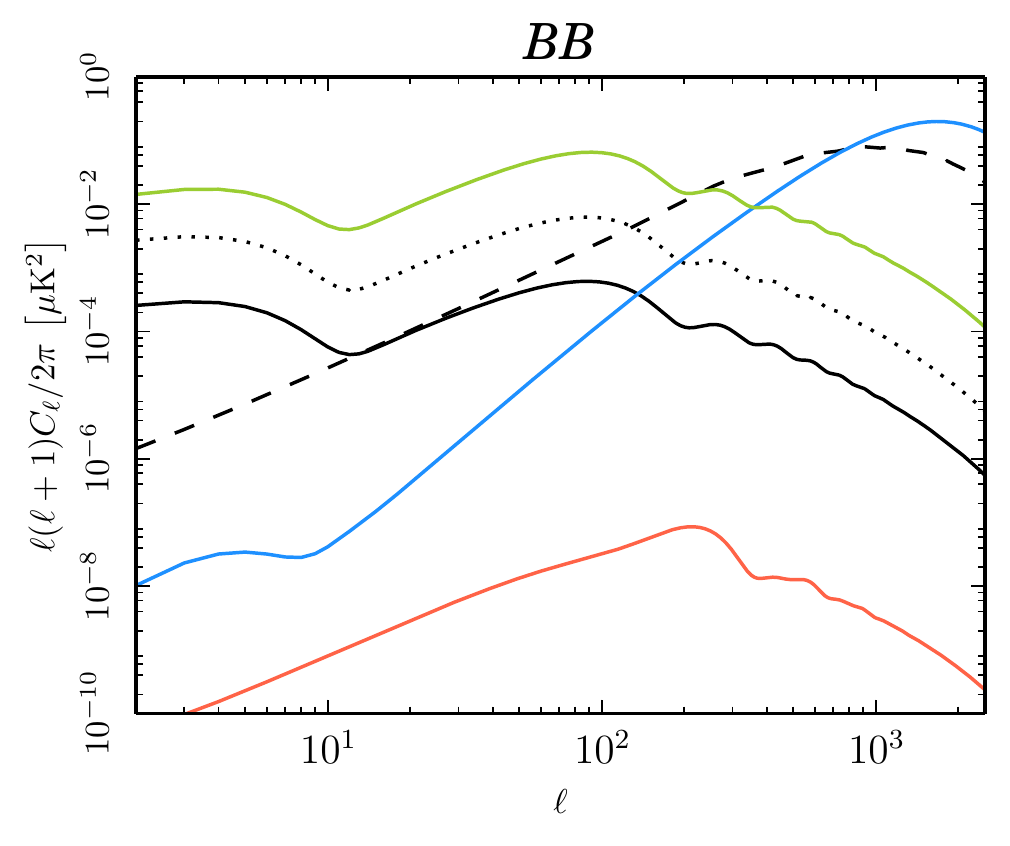}
\caption{\footnotesize\label{fig:Spettri} Magnetically-induced CMB anisotropies angular power spectra in temperature [upper left], E-mode polarization [upper right], T-E cross correlation [lower left]. The colors represent: compensated scalar-pink, vector-blue, tensor-orange, passive scalar-yellow, tensor-green; black is the primary CMB. The B-mode polarization [lower right] color code is analogous but in this case black solid and dotted lines represent primary tensor modes for tensor to scalar ratios of r=0.1 and 0.01 and the dashed line represents the lensing B-mode from primary scalar perturbations. The field amplitude is 3 nG and the spectral index is the almost scale-invariant.}
\end{figure*}

\begin{figure*}[!htb]
\includegraphics[width=0.5\columnwidth]{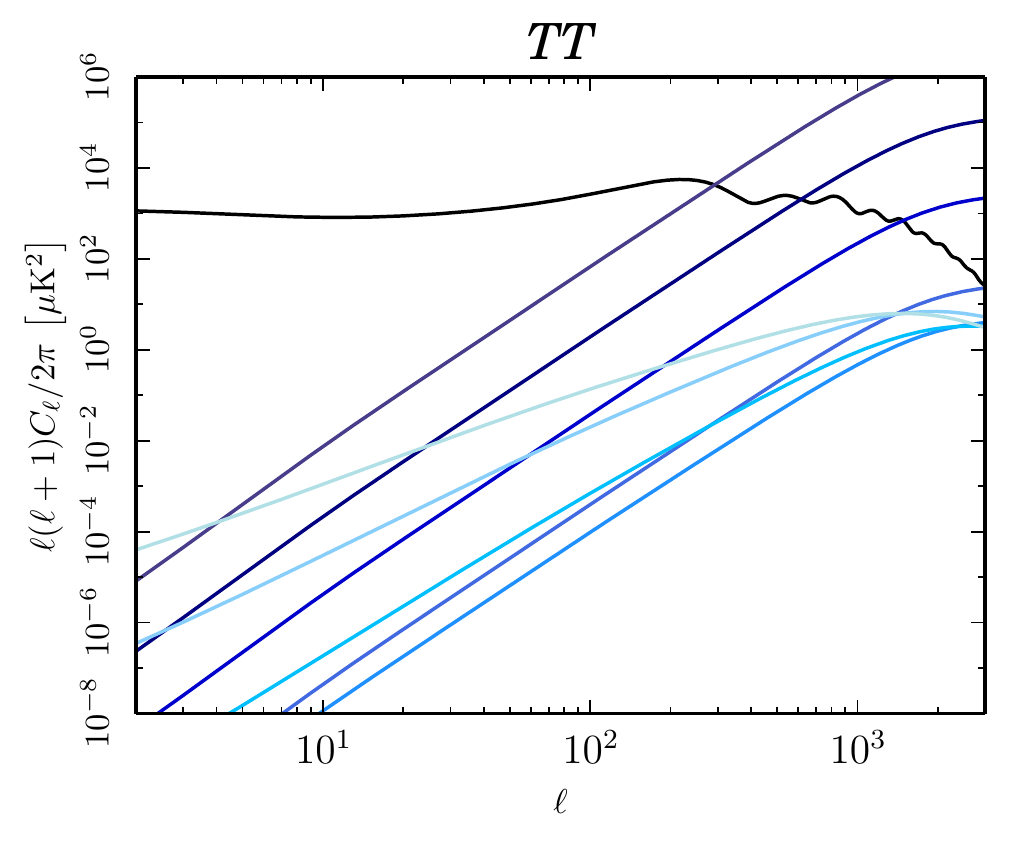}
\includegraphics[width=0.5\columnwidth]{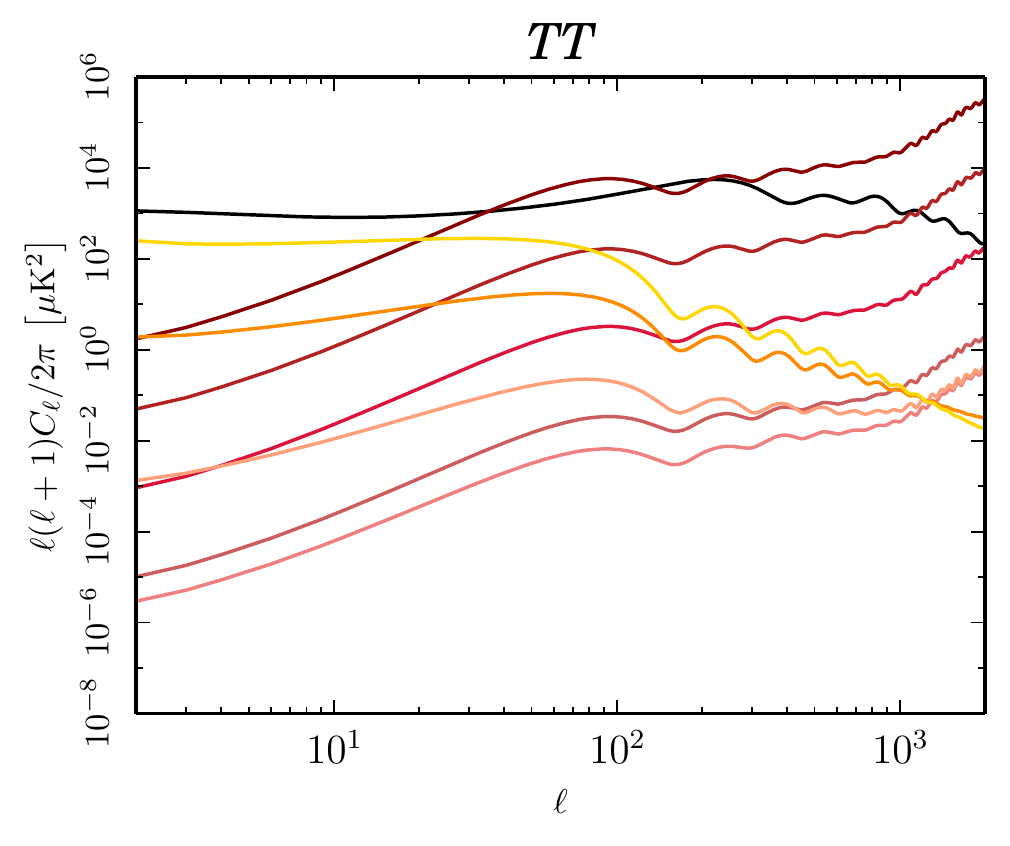}\\
\caption{\footnotesize\label{spectral} Magnetically-induced CMB anisotropies angular power spectra in temperature for vector [left] and passive tensor [right]. The colors represent the different spectral indices, darker corresponds to bigger indices, starting from 2, whereas lighter represent negative indices following the shading, ending in quasi scale invariant.}
\end{figure*}

We use an upgrade, described in the next section, of the camb code \cite{Lewis:1999bs} extension that includes all PMFs gravitational contributions developed in \cite{Finelli:2008xh,Paoletti:2008ck,Paoletti:2010rx,Paoletti:2012bb,Ade:2015cva} to predict the angular power spectra of magnetically-induced perturbations. In Fig. \ref{fig:Spettri} we show the different modes in different colours for a value of the field amplitude and spectral index of 3 nGauss and almost scale invariant spectrum ($n_{\mathrm B} =-2.9$). The dominant contributions are given by the vector and passive scalar modes on small angular scales together with the passive tensor mode on large angular scales. The latter is particularly relevant for B-mode polarization providing a signal which is somehow similar to the one by primordial gravitational waves at least for almost scale invariant PMFs. In the perspective of a future detection of B-mode polarization on large angular scales the consideration of this additional possible contribution will be necessary to use the detected signal for the study of the inflationary epoch and avoid confusion.  
The shape of the magnetic contributions to the angular power spectrum reflects the behaviour of the source terms and in particular s shown in \cite{Finelli:2008xh,Paoletti:2008ck} it is a white noise for indices $n_{\mathrm B} >-3/2$ and infrared dominated as $k^{2n_{\mathrm B} +3}$  for lower indices. This behaviour is presented in Fig. \ref{spectral} where are shown the temperature angular power spectra of vector and passive tensor modes for different spectral indices.
\section{Methodology updates}
The different gravitational contributions of PMFs on the Einstein-Boltzmann equations have been implemented into camb  \cite{Lewis:1999bs} and \texttt{cosmomc}  \cite{Lewis:2002ah} \footnote{Two other different implementations based on similar assumptions but different numerical settings are done in \cite{Shaw:2009nf,Shaw:2010ea,Zucca:2016iur}.}  and have been used in our previous analyses \cite{Finelli:2008xh,Paoletti:2008ck,Paoletti:2010rx,Paoletti:2012bb,Ade:2015cva}. The magnetic part of the code is based on numerical fits to the large scale part (the only relevant for CMB data) of the analytic solutions to the convolution integrals in Eq.\ref{tensor} \cite{Finelli:2008xh,Paoletti:2008ck} that in their original form involve a large number of hypergeometric functions. The previous fits were computed using a grid for the full spectra expressions and were derived by fitting sets of parameters for each grid element. The fits used in the current work have been upgraded with a new approach which introduces a multiparameter sampling not of the full formula, like in the previous ones, but for each coefficient of the expression which is then fitted for the involved parameters.  This new method allows to correctly sample the regions near the multiple poles of the coefficients of the expressions that in the previous analyses had a lower accuracy because of the full single formula parametric fit. We have increased the number of fits computed reducing the spectral indices grid spacing for the sampling, increasing the accuracy over the whole range explored. Applying this new sampling technique the new analytical fits used in this work represent the analytic solutions to the convolution integrals with a precison of one part over $10^4$ for negative indices and $10^5$ for positive indices. The new fits have also lead to an increased numerical stability which adds to the generally improved numerical accuracy of the 2018 camb version. This new precision has allowed to increase the range of multipoles considered for the passive tensor magnetically-induced mode to $\ell=2000$ (the previous code used 400 as a conservative maximum multipole) which is the maximum multipole where we trust the numerics of our modified version of camb. Indeed, tensor magnetically induced modes do not decay once they enter in the Hubble radius since thery are driven by the magnetic anisotropic stress. Therefore their signal remains high compared to the standard primary tensor modes. This improved accuracy is particularly relevant for the constrains on the PMFs, especially when considering the forecasts for future experiments that include high precision small scales data.  
In addition to the improvements described in the following analysis we consider all the dominant contribution including the passive tensor and scalar modes, with the latter one not considered in previous analyses \cite{Paoletti:2010rx,Paoletti:2012bb,Ade:2015cva}. 
    
\section{Constraints with CMB data}
We perform a Bayesian analysis deriving the probability distributions of cosmological parameters plus the two parameters characterizing the PMFs configuration, i.e. the amplitude on the Mpc scale and the spectral index\footnote{In case of real data likelihood we add also the nuisance's parameters}. We perform the analysis through a dedicated extension of the Markov Chain MonteCarlo sampler \texttt{cosmomc}  \cite{Lewis:2002ah}.
We vary the standard cosmological parameters: baryon density $\omega_{b}=\Omega_{b} h^2$, the cold dark matter density 
$\omega_{c}= \Omega_{c}h^2$ , the reionization optical depth $\tau$ ,  the acoustic scale $\theta$, and the primordial power spectrum amplitude  $\ln ( 10^{10} A_S )$ and spectral index $n_S$. The magnetic parameters are indicated by  $B_{1 {\mathrm{Mpc}}}$ and $n_{\mathrm B}$ and are varied with flat priors [0,10] nG and [-2.9,3] respectively. 
We restrict our anaylsis to three massless neutrinos, since the small neutrino mass allowed by current observations leads to a small effect  \cite{Shaw:2010ea} without inserting additional degeneracies.
%
%as shown in \cite{Shaw:2010ea} massive neutrinos modifies the large angular scale part of vector and scalar modes leading to a negligible effect on the PMFs amplitude constraints, and as for possible study of joint analysis in the perspective of future experiments which will be highly sensitive to both neutrinos and PMFs we must consider that the really high sensitivity to neutrinos characteristics poses the problem of the choice of particle physics assumptions like the neutrino hierarchy, especially in high sensitivity and resolution conditions see for example \cite{DiValentino:2016foa}, for this reason for both real data and forecasts we choose to use massless neutrinos.
We sample the posterior using the Metropolis-Hastings algorithm \cite{Hastings:1970aa} imposing the Gelman-Rubin convergence criterion to \cite{Gelman:1992zz} of $R-1 < 0.01$. We consider lensing effect on only primary anisotropies, since modelling the contribution of PMFs to the large scale structures would require non-linear MHD treatments currently not available in the context of Einstein-Boltzmann codes. We therefore neglect the impact of PMFs on the lensing effect and viceversa the lensing of magnetically induced angular power spectra. 
\subsection{Planck 2018}
\begin{figure*}[!htb]
\includegraphics[width=1\columnwidth]{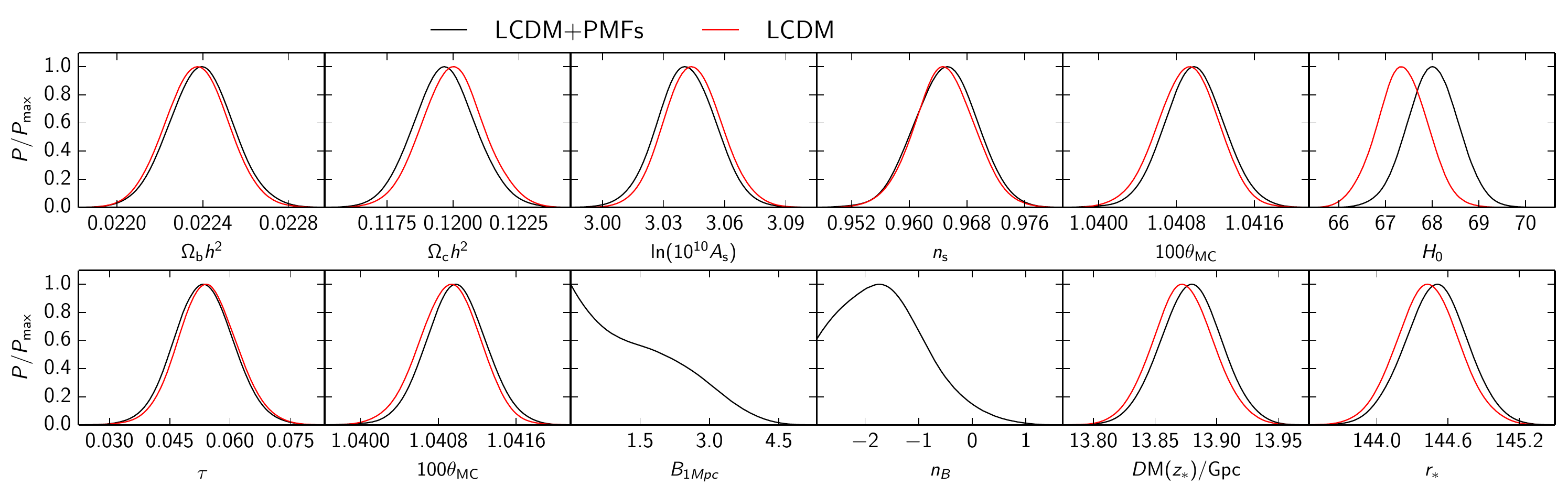}
\caption{\footnotesize\label{1D} Marginalized posterior of the Planck 2018 results compared with the standard cosmological model.}
\end{figure*}

We first analyse the Planck 2018 data release using the combination of the high multipole baseline likelihood in temperature and polarization (Plik TT,TE,EE), the low-multipole Commander likelihood in temperature (lowl)  and the new cross spectra 100x143 GHz based likelihood in polarization (LowE) based on the first release of the High Frequency Instrument maps with the SROLL algorithm \cite{Aghanim:2019ame,Aghanim:2018fcm}. To complete the Planck baseline likelihood, in addition to the primary anisotropies we include also the Planck lensing likelihood \cite{Aghanim:2018oex}.
The baseline Plik TT,TE,EE+ lowl+LowE+lensing lensing provides a 95\% C.L. on the amplitude of the fields on the Mpc scale: $B_{1 \,\mathrm{Mpc}}<3.4$ nG and marginalized 95\% C.L. $n_{\mathrm B}<-0.21$. We stress that the marginalized posterior probability for the spectral index only reflects the lower amplitude allowed for positive spectral indices compared to negative ones.
Figure \ref{1D} shows the comparison of the marginalized posteriors for parameters including PMFs with the standard $\mathrm{\Lambda}$CDM. We note how the presence of PMFs does not significantly affect the standard $\mathrm{\Lambda}$CDM parameters: the induced shift in the Hubble parameter is simply given by the assumption of no neutrino mass with respect to the case of the minimal neutrino mass of Planck baseline results for $\mathrm{\Lambda}$CDM.  
 %We note a significant enahancement in the marginalized value of the Hubble constant $H_0=68.01\pm 0.54$ with respect to the standard LCDM case (best fit likelihoods 1391.01 for the LCDM and 1389.86 for the PMFs). The increase in the value is given by the impact of the Lorentz force on baryons and the additional anisotropic pressure provided by the PMFs that modifies the dynamic of recombination and affects the expansion rate (as also shown by the constraints on PMFs amplitude from Big Bang Nucleosynthesis).
\subsection{Planck 2018 results in combination with ground based experiments}
We now complement  Planck 2018 with B-mode data from ground based high sensitivity polarization experiments such as BICEP/Keck array \cite{Ade:2018iql} and South Pole Telescope \cite{Keisler:2015hfa}. 

The inclusion of the B-mode polarization on large and intermediate angular scales from BICEP/Keck 2015 data (hence BK15) modifies the posterior marginalized probability distribution allowing for a slighter larger signal from PMFs. The limit in this case is $B_{1 \,{\mathrm Mpc}}<3.5$ nG at 95\% C.L slightly larger than Planck 2018 alone. 

\begin{figure*}[!htb]
\includegraphics[width=\columnwidth]{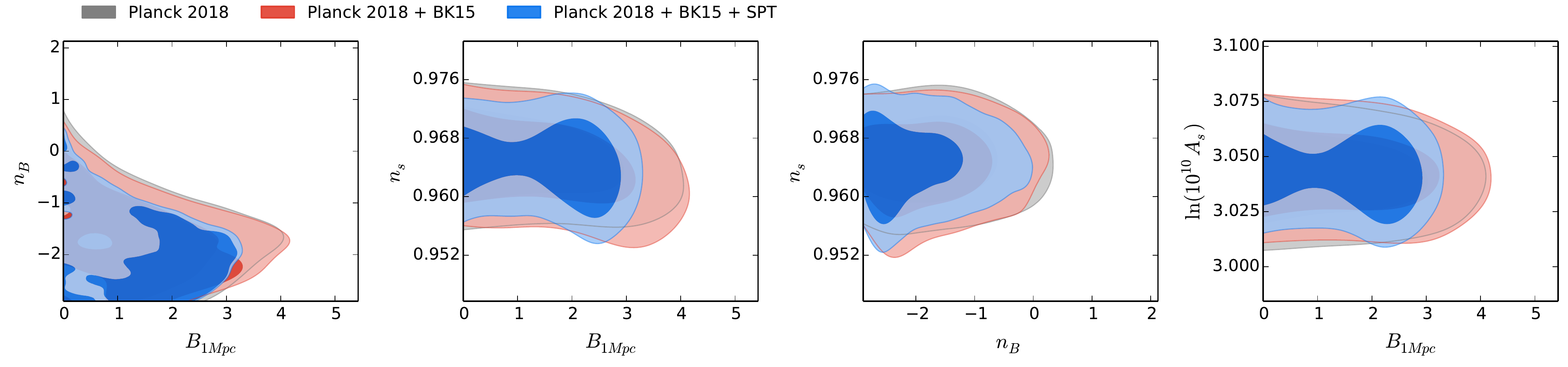}
\caption{\footnotesize\label{V1} Two-dimensional marginalized posterior distributions for the magnetic parameters, first panel, and the comparison of the PMFs amplitude with some of the standard parameters, the scalar spectral index, the $\sigma_8$ and finally the Hubble constant. }
\end{figure*}

We then further combine Planck 2018 and BK15 with the B-mode measurement on small angular scales provided by the South Pole Telescope \cite{Keisler:2015hfa} in order to increase the multipole range in the B-mode data. We use the likelihood and data of SPT for cosmomc provided in \cite{Zucca:2016iur}.  Since we combine the two datasets from BK15 and SPT in the B-mode polarization we cut out overlapping multipoles in order to avoid considering possible cross correlations between the two datasets. 
For this reason we cut the lowest multipoles of the SPT dataset to $\ell$ of 500.
In Fig.\ref{V1} we show the correlation of the magnetic field amplitude with the spectral index and a set of standard parameters and in particular the scalar spectral index and the amplitude of scalar perturbations. Beside the degeneracy between the amplitude and the spectral index we do not have any relevant degeneracy with the standard cosmological parameters. In Table \ref{Table1} we present the results for the magnetic parameters.

\begin{table}[tbp]
\centering
\begin{tabular}{|l|c|c|c|}
\hline
Parameter/Data& Planck 2018 &Planck 2018+BK15 & Planck 2018 + BK15+SPT Pol\\
\hline
$B_{1 \,\mathrm{Mpc}}$ [nG] & <3.4 & <3.5& <2.8\\
$n_{\mathrm B}$ & <-0.2 & <-0.3&<-0.6\\
\hline
\end{tabular}
\caption{\label{Table1} Constraints on the PMFs amplitude with CMB data marginalizing over the spectral index}
\end{table}
As in previous works \cite{Ade:2015cva} we complement the previous results obtained allowing the spectral index to vary with the constraints on the amplitude for physical fixed values for the spectral index.
%Considering that there is no-detection of PMFs amplitude but only upper limits, adding the high degree correlation of the amplitude with the spectral index, as we already mentioned the constraints on the spectral index are purely indications with no statistical significance. Therefore, in order to explore the possible implications of the CMB data on the PMFs configuration, and as a consequence the implications for possible generation mechanisms, we explore a couple of specific cases where the spectral index is fixed. 
In particular we focus on the minumum allowed spectral index for causally-generated post-inflationary fields, the case $n_{\mathrm B}=2$ \cite{Durrer:2003ja}: for this case Planck combined with BICEP/KECK gives  $B_{1 \,\mathrm{Mpc}}<6$ pG at 95\% C.L., when adding also the high multipoles B-mode from the South Pole Telescope we obtain $B_{1 \,\mathrm{Mpc}}<3$ pG at 95\% C.L.. The second case we consider is the almost scale invariant  PMFs configuration $n_{\mathrm B}=-2.9$, this represents possible PMFs generated during inflation, although inflationary generation mechanisms are not limited to infrared indices but span the entire range from negative to positive depending on the specific type of mechanism considered. This case is particularly interesting because of the shape of the tensor passive magnetically-induced mode, as shown in Fig.\ref{fig:Spettri}, this is slightly similar to the angular power spectrum of primary tensor modes and may represent a nuisance for the detection for primordial gravitational waves from inflation. In this case the bound on the amplitude of the fields is given by $B_{1 \,\mathrm{Mpc}}<2.0$ nG at 95\% C.L. for Planck and BICEP/KECK and $B_{1 \,\mathrm{Mpc}}=1.5_{-1.4}^{+0.9} $ nG when adding the SPT data, with a marginal detection which is not confirmed at 99\% C.L. $B_{1 \,\mathrm{Mpc}}<2.5$ nG.
For the full combination of Planck BICEP/KECK and SPT we also analised the case $n_{\mathrm B}=0$ which provides  $B_{1 \,\mathrm{Mpc}}<0.3$ nG at 95\% C.L..\\

\begin{figure*}[!htb]
\includegraphics[width=0.75\columnwidth]{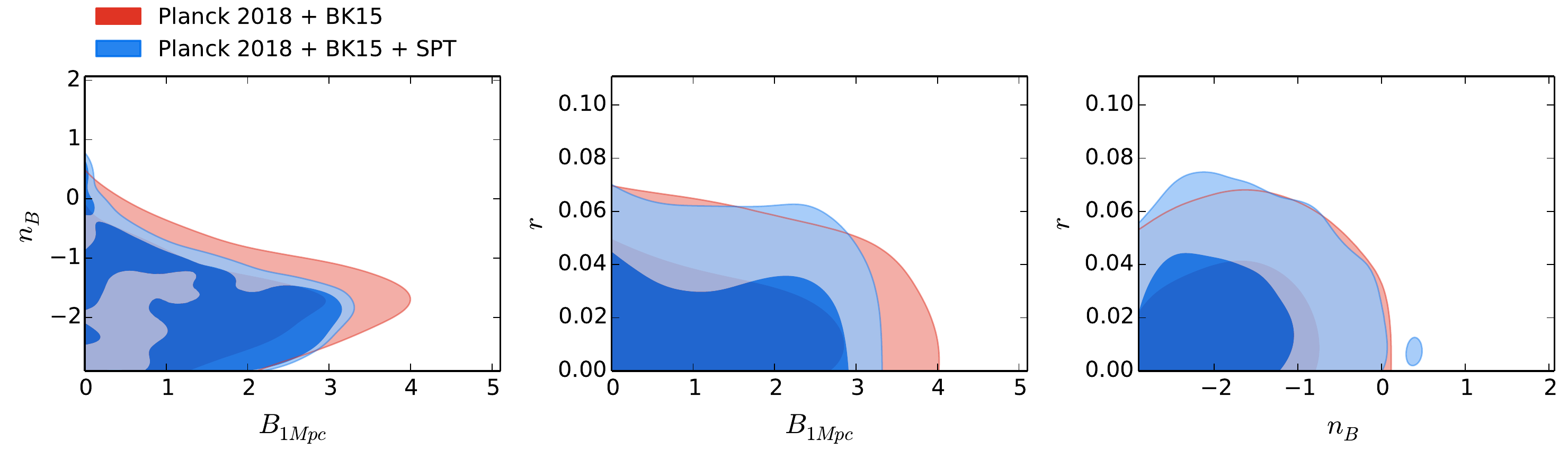}\includegraphics[width=0.25\columnwidth]{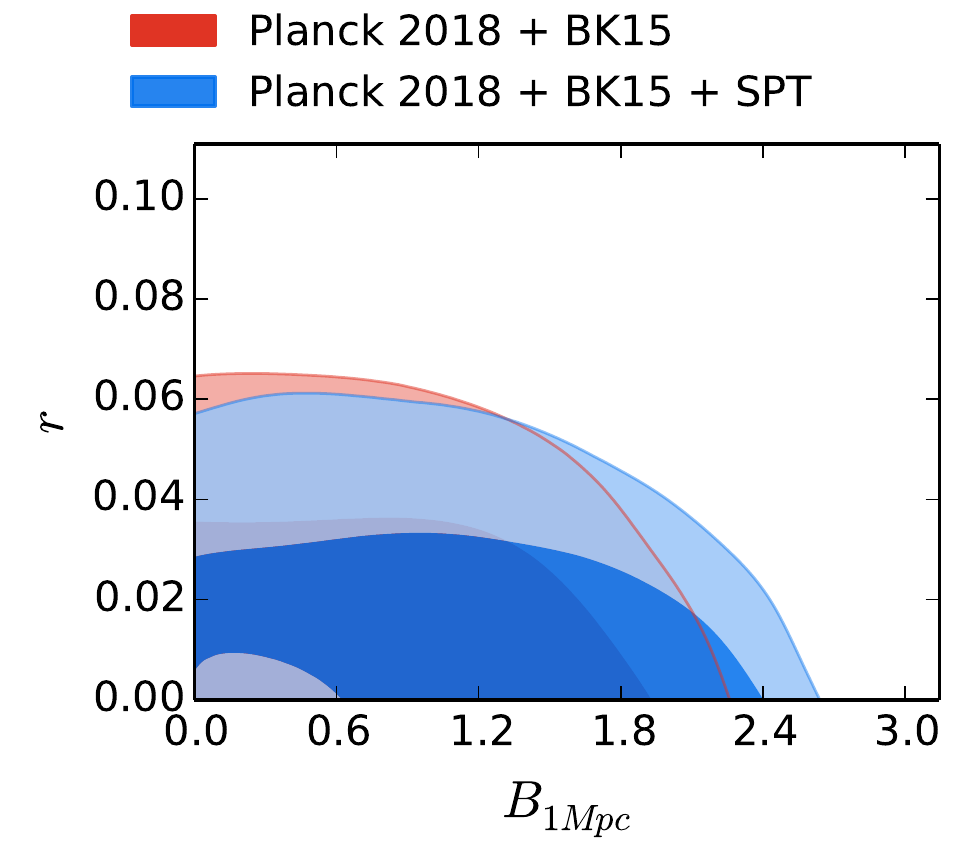}
\caption{\footnotesize\label{r1} Contour plot of the magnetic fields amplitude and spectral index on the left, the correlation between the PMFs amplitude and the tensor to scalar ratio in the middle and with the spectral index on the right. Far right panel shows the correlation between the fields amplitude and the tensor to scalar ratio for the almost scale-invariant case. }
\end{figure*}

\begin{figure*}[!htb]
\includegraphics[width=0.35\columnwidth]{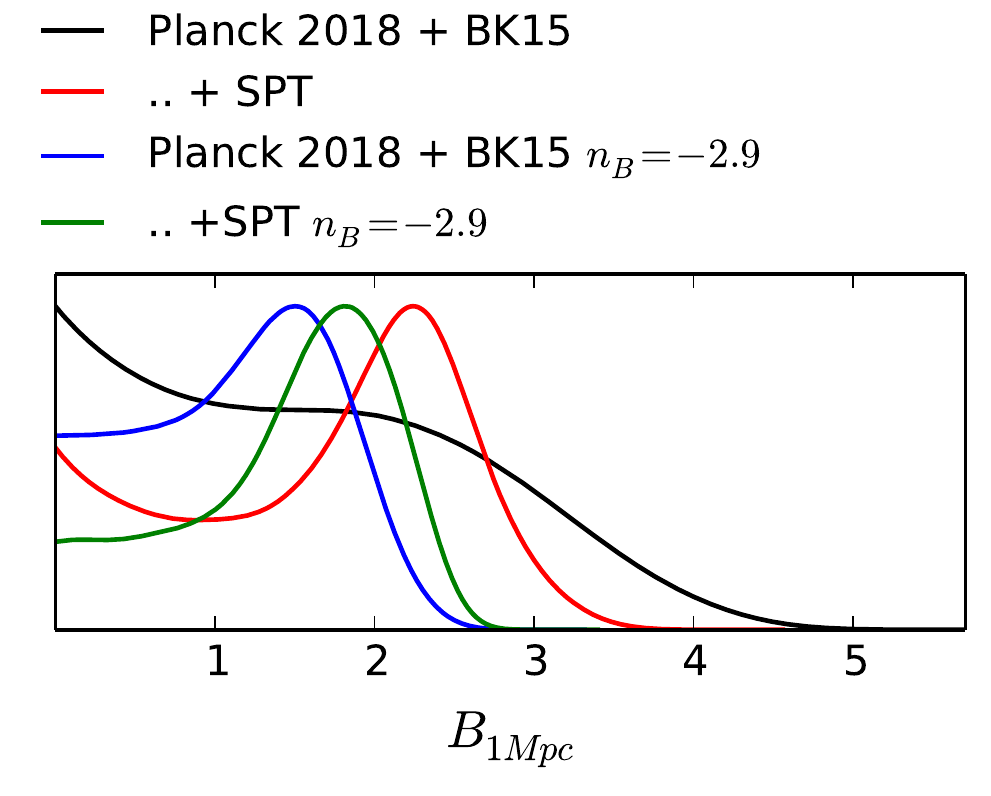}
\includegraphics[width=0.72\columnwidth]{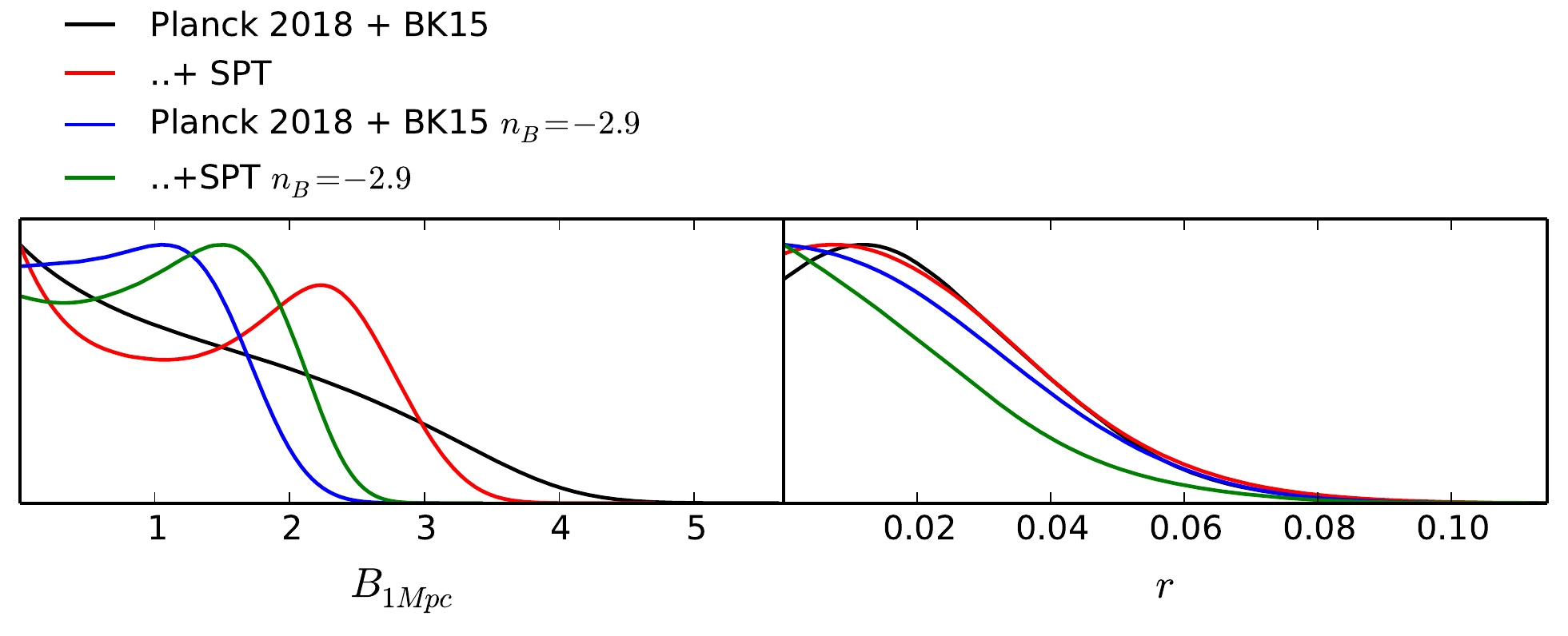}
\caption{\footnotesize\label{pippo} Marginalized posterior distribution of the magnetic fields amplitude for different data combination and settings of the field for the LCDM+PMF case on the left and jointly with the tensor to scalar ratio for LCDM+r+PMF on the right.}
\end{figure*}

Considering the possible contamination of the primordial gravitational wave signal by the tensor passive contribution we explore the case where together with the magnetically induced tensor mode we have also a primary contribution, the tensor to scalar ratio is allowed to vary together with the magnetic fields amplitude. In Fig.\ref{r1} we show the two-dimensional probability distribution for the PMFs parameters and the tensor to scalar ratio for the case with the combination Planck and BICEP/KECK and with the addition of SPT. The constraints on the two amplitudes are $B_{1 \,\mathrm{Mpc}}<3.4$ nG and $r<0.06$ at 95\% C.L. and $B_{1 \,\mathrm{Mpc}}<2.9$ nG and $r<0.06$ at 95\% C.L., we note that the case marginalized on the spectral index of the PMFs does not show a significant correlation between the magnetic parameters and the tensor to scalar ratio. We perform the same analysis for the almost scale invariant case whose spectral shape is limited to be very similar to the primary tensor modes. The constraints on the two amplitudes for the cases without and with SPT are $B_{1 \,\mathrm{Mpc}}<1.8$ nG and $r<0.06$ at 95\% C.L. and $B_{1 \,\mathrm{Mpc}}<2.2$ nG and $r<0.055$ at 95\% C.L.. In the scale invariant case there is a slight correlation between the amplitude and the tensor to scalar ratio but for the current instruments sensitivity is not relevant for the results, the tensor to scalar ratio limit is almost unchanged with respect to results without the PMFs contribution.
In Fig.\ref{r1} and Fig.\ref{pippo} we show a summary of the different constraints.

\section{Forecasts for future experiments}
After presenting the Planck 2018 update results alone and in combination with ground based experiments we proceed with the forecasts for future experiments. 
%We consider the LiteBIRD mission alone and in combination with future ground based experiments Simons Observatory and CMB Stage IV. 
\subsection{LiteBIRD}

We consider the LiteBIRD mission \footnote{http://litebird.jp/eng/} \cite{Hazumi:2019lys} 
among the various proposals for a fourth generation space mission dedicated to CMB anisotropy measurements \cite{Hazumi:2019lys,Hanany:2019lle,Delabrouille:2019thj}.
%the LiteBIRD mission\footnote{http://litebird.jp/eng/} \cite{Hazumi:2019lys}.
LiteBIRD is a proposal for a satellite downselected by the Japanese space agency JAXA, 
with the contributions of USA, Europe and Canada, with planned launch in 2027. 
Its main goal is the study and characterization of the CMB polarization at large and intemediate angular scales. It has fifteen frequency channels spanning from 40 GHz to 402 GHz, a range optimized for the foreground removal and it will have a payload-focal plane design to provide very high sentivity measurements of the E and B-mode polarization up to a multipole of several hundreds.
We summarize the instrumental characteristics of the central frequency channels we used for the analyses in Table \ref{LiteInstr}  \cite{Hazumi:2019lys} which updates previous configurations \cite{Matsumura:2016sri, Finelli:2016cyd}. We consider the highest and lowest frequency channels as foreground tracers and we assume they will be used to clean the central ones. We assume only instrumental noise and ideal foreground cleaning with a inverse-Wishart likelihood on mock datasets \cite{Finelli:2016cyd}. This choice represents a very optimistic setting and therefore provide a snapshot of the limits on primordial magnetic fields which can be reached by future experiments.  We consider the lensing signal as additional noise and simply add a lensing fiducial, derived with the same cosmological parameters as the primary CMB fiducial, to the noise bias. We will also investigate the impact on the results of delensing possibilities.
 
\begin{table}[tbp]
\centering
\begin{tabular}{|c|c|c|c|c|}
\hline
Frequency GHz & FHWM [arcmin] & T Sensibility[$\mu K$ arcmin] & P Sensibility[$\mu K$ arcmin] \\
\hline
$78$  & 39 &9.56 & 13.5 \\
$89$ & 35 & 8.27&11.7\\
$100$ & 29 &6.50 &9.2\\
$119$ & 25 &5.37 &7.6\\
$140$ & 23 &4.17 &5.9\\
$166$ & 21 & 4.60&6.5\\
$195$ & 20 & 4.10&5.8\\
\hline
\end{tabular}
\caption{\label{LiteInstr} LiteBIRD instrumental characteristics \cite{Hazumi:2019lys}}
\end{table}
In Table \ref{LiteBIRD} we present different cases for the LiteBIRD study. 
Being cosmic variance limited in the E-mode polarization we would like to investigate the LiteBIRD capabilities to constraint magnetic fields amplitude using only temperature and E-modes. When only temperature and E-mode polarization are included we obtain $B_{1 \,\mathrm{Mpc}}<3.8 $ nG at 95\% C.L... This limit is comparable to what obtained by Planck and shows the capabilities of the addition of cosmic variance limit E-mode polarization measure for magnetic fields, eventhough the impact of PMF on E-mode is much lower than in T and B-mode, the high sensitivity of LiteBIRD is already enough by itself to reach a Planck-like limit. 

\begin{figure*}[!htb]
\includegraphics[width=\columnwidth]{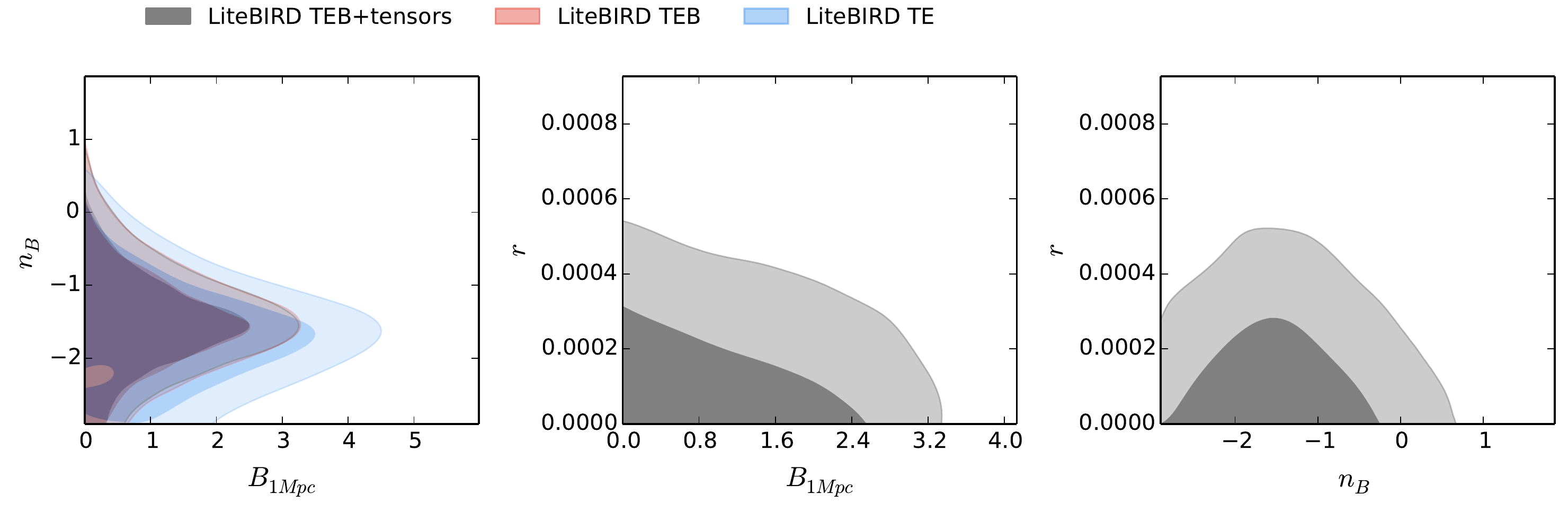}
\caption{\footnotesize\label{LiteBIRDAlone} Marginalized posterior distribution of the magnetic fields amplitude for different data combination and settings of the field for the LCDM+PMF case, together with the tensor to scalar ratio for LCDM+r+PMF on the right. Note that red and grey curves in the left panel are overimposed since the results do not change sensibly adding the tensor to scalar ratio.}
\end{figure*}

We now add the B-mode channel which significantly reduces the bound on the fields amplitude thanks to the very high sensitivity measurement: $B_{1 \,\mathrm{Mpc}}<2.7 $ nG at 95\% C. L.. 
As done for real data we also test the case when primary tensor modes from inflation are added. We vary the tensor to scalar ratio together with LCDM and magnetic parameters assuming it vanishing in the fiducial angular power spectra.  We obtain $B_{1 \,\mathrm{Mpc}}<2.7 $ nG at 95\% C.L. equal to the case with only PMFs but the result does not show any strong degeneracy with the primordial tensor modes as shown by Fig.\ref{LiteBIRDAlone}. The limit on r is $r<0.0004$ compatible with the foreground and magnetic free LiteBIRD predictions. 
When we limit our analysis to specific configurations of PMFs, again to the lower bound for causally-generated field $n_{\mathrm B}=2$ and the almost scale invariant case $n_{B}\sim -2.9$ we obtain respectively $B_{1 \,\mathrm{Mpc}}<4$ pG  and $B_{1 \,\mathrm{Mpc}}<0.6$ nG at 95\% C.L. strongly reducing the allowed amplitude for the almost scale invariant case with respect to the current limit derived using the combination of large and small scales experiments.

\begin{table}[tbp]
\centering
\begin{tabular}{|l|c|c|c|}
\hline
Parameter/Data& LiteBIRD TE &LiteBIRD TEB & LiteBIRD TEB + primordial tensors\\
\hline
$B_{1 \,\mathrm{Mpc}}$ [nG] & <3.8 & <2.7& <2.7\\
$n_{\mathrm B}$ & <-0.26 & <-0.19&<-0.16\\
\hline
\end{tabular}
\caption{\label{LiteBIRD} Constraints on the PMFs amplitude with CMB data marginalizing over the spectral index}
\end{table}

These results show how the LiteBIRD  sensitivity will lead to constraints by T and E only at the level of those obtained by Planck data. By adding the contribution of B-mode polarization further 
tightens the constraints by LiteBIRD without any significant degeneracy with primary tensor modes from inflation.
%We stress again that we provide constraints on the spectral index as a purely indication of the stronger constraints on the amplitude for positive indices, we remark that in the absence of a detection the constrain on the spectral index is strongly prior dominated.

\subsection{Combination of LiteBIRD and future ground based experiments}
The launch of LiteBIRD is planned for 2027 and will carry out the fourth generation of CMB observations from space. In the timescale leading to the launch and also during the LiteBIRD observational time advancements from ground based experiments are expected. In particular, we refer to the future ground based experiments Simons Obsevatory \cite{Abitbol:2019nhf, Ade:2018sbj} and CMB Stage IV \cite{Abazajian:2019eic,Abazajian:2016yjj}. The enviroment offered from the ground allows for very high sensitivity observations thanks to the possibility of large focal planes containing tens of thousands of receiver and the possibility to observe in great depth small region of the sky. On the other side ground based observatories due to their nature suffer from two limitations, namely, sky coverage and frequency range. The former is due to a limited sky coverage and therefore the impossibility to cover the largest angular scales, where most of the primary inflationary B-mode signal resides, and a larger sampling variance on the intermediate scales. The latter is due to the atmosphere which is opaque for frequencies larger than 300 GHz limiting the higher ``foregrounds monitor'' frequencies which are crucial for dust and cosmic infrared background contaminations. 
The contemporaneity of a space borne experiment dedicated to the large angular scales and ground based observatories on small angular scales represents a unique opportunity to exploit the potential of CMB polarization in the context of cosmic magnetism. For this reason we investigate in the following subsections the case of the combination of LiteBIRD with ground based future experiments.
In order mitigate effects due to the cross correlation between maps we consider LiteBIRD up to a multipole of 600 and ground based experiments for higher multipoles. At multipoles around 600 the signal to noise of LiteBIRD and ground based experiments are comparable.
 
\subsubsection{Combination with Simons Observatory}

We study the combined analysis of LiteBIRD and the Simons Observatory that represents the near future of ground based experiments. For Simons Observatory we consider the central frequency channels as in Table.\ref{Tableso} with a wide survey-like sky fraction of 40\%.  We analise several possible PMFs configurations for the combination LiteBIRD+SO. 
The standard case in which the PMFs are marginalized over the spectral index and no primordial tensor modes are considered provides  $B_{1 \,\mathrm{Mpc}}<1.0$ nG with $n_{\mathrm B}<-0.6$. We note a significant improvement with respect to current results reported in previous section. Another aspect that we to stress is that the polarization sensitivity is high enough to provide constraints even without considering the temperature. This aspect is particularly promising because the temperature signal is contaminated on small angular scales by astrophysical signals and secondary anisotropies, and has been shown in \cite{Paoletti:2012bb,Ade:2015cva} that these may be correlated with the PMFs contributions and lead to contamination of the PMFs limits. Therefore we tested a case where we exclude the temperature from the analysis using only E-mode, the T-E cross correlation and B-mode polarization obtaining $B_{1 \,\mathrm{Mpc}}< 1.1$ nG, at the same level as considering the temperature. Considering that the contamination from foregrounds on the small angular scales in polarization is almost negligible, we can conclude that in the future the large scale contamination by polarized dust and synchrotron, and possibly new signals still to discover, will be the most serious issue to deal with.

\begin{figure*}[!htb]
\includegraphics[width=\columnwidth]{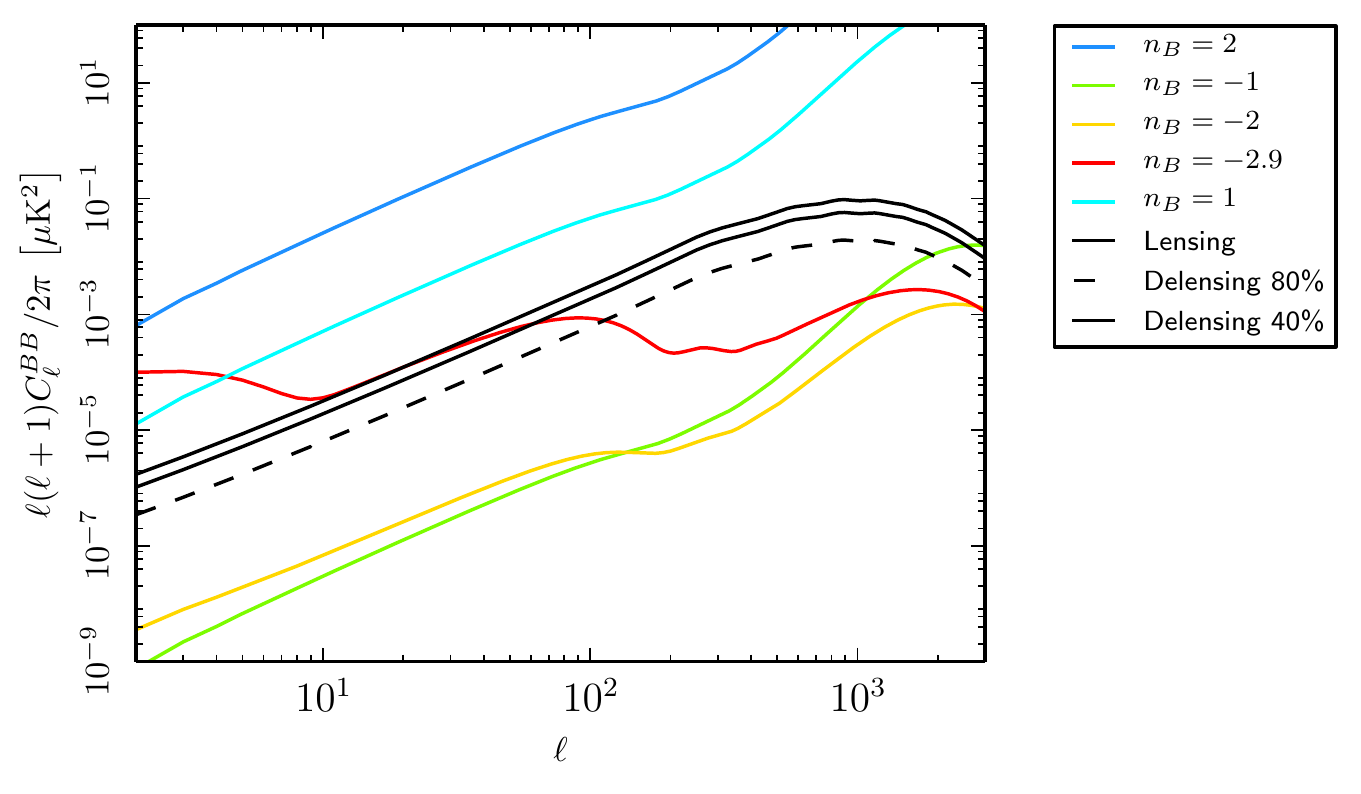}
\caption{\footnotesize\label{Dele} B-mode angular power spectrum for the magnetically-induced modes compared with delensing signal.}
\end{figure*}

\begin{figure*}[!htb]
\includegraphics[width=0.33\columnwidth]{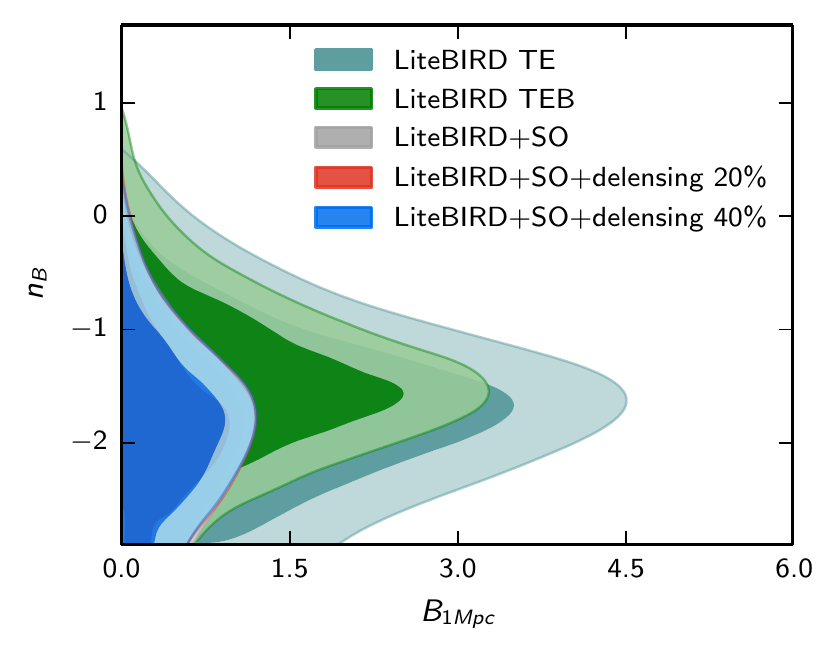}\includegraphics[width=0.33\columnwidth]{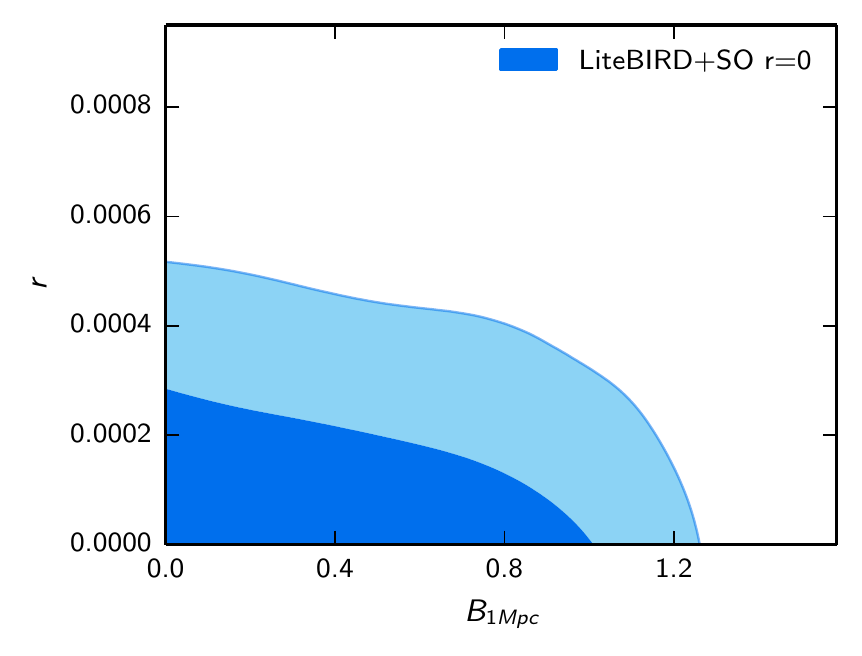}\includegraphics[width=0.33\columnwidth]{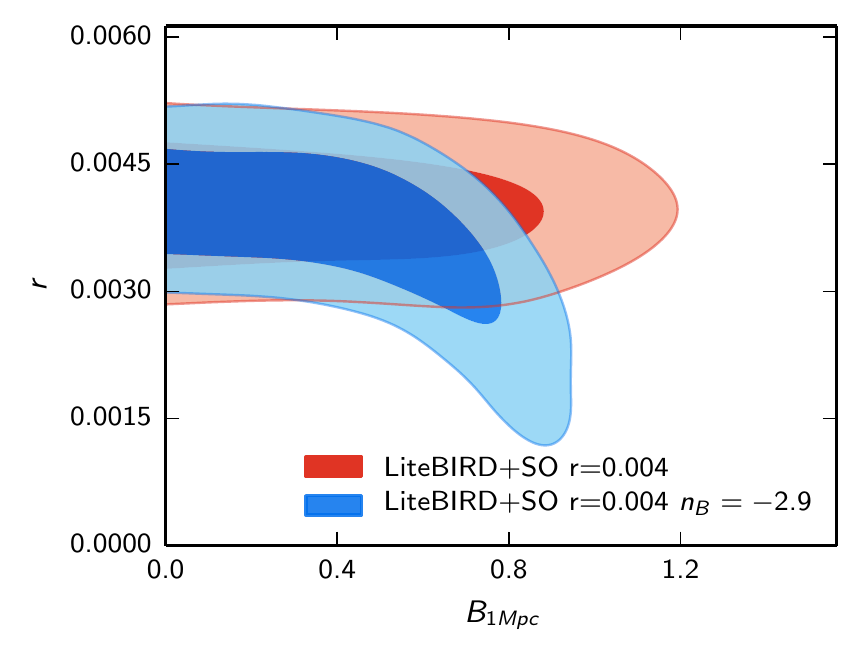}
\caption{\footnotesize\label{plotFuturo} Contour plot of the magnetic fields amplitude and spectral index on the left and tensor to scalar ratio on the right. Left panel shows the comparison between Planck 2018+BICEP KECK and the combination with SPT. The middle panel considers the case with a null tensor to scalar ratio whereas the right panel the case where it is assumed an r=0.004}
\end{figure*}

If we assume a perfect cleaning of foregrounds like in our semi-idealistic treatment, the lensing signal represents the only parasitic signal for B-mode polarization. 
Future experiments will help with delensing either by iterative algorithms of CMB polarization or with the cross correlation with tracers of large scale structure as the Cosmic Infrared Background \cite{Errard:2015cxa}.  We consider two possible options for delensing of both LiteBIRD and SO, both possibilities are conservative: delens up to the 20\% and 40\% of the signal. Both the cases with 20\% and 40\% delensing give  $B_{1 \,\mathrm{Mpc}}<1 $ nG the same results as the fully lensed case as shown also in the first panel of Fig.\ref{plotFuturo} where we present the two dimensional contours for the LiteBIRD+SO data combination in comparison with LiteBIRD alone. 
The ineffectiveness of the delensing is due to the shape of the contribution of the PMFs to the B-mode angular power spectrum. The vector mode peak in the B-mode spectrum is slightly offset with respect to the lensing and for a wide range of spectral index the signal around the peak is larger than the lensing one leaving possible improvements to the intermediate multipole region. this region due to the steepness of the vector signal can only marginal benefit the low tail of the spectral indices range as shown in Fig. \ref{Dele}.

In fact we tested also an extreme value for the delensing that assume to be able to delens up to the 80\% of the sky with the PMFs configuration with marginalization over the spectral index obtaining the same results of 1 nG upper limit. 
 
The situation changes when we consider specific PMFs configuration, in particular for the almost scale invariant case without delensing we obtain $B_{1 \,\mathrm{Mpc}}<0.55 $ nG at 95\% CL which represents a good improvement with respect to current data mainly provided by the large scale impact of the almost scale invariant configuration. If we consider the delensing for this specific case as described before we obtain $B_{1 \,\mathrm{Mpc}}<0.50$ nG and $B_{1 \,\mathrm{Mpc}}<0.44 $ nG at 95\% CL for respectively 40\% and 80\% delensing showing that for the infrared tilted spectra the delensing becomes effective in reducing the upper limit.

For PMFs configuration which allows a generation also through the causal mechanism channel, $n_{\mathrm B}=2$ we get  $B_{1 \,\mathrm{Mpc}}< 0.5$ pG at 95\% CL, an order of magnitude improvement with respect to the current data provided by the small scales sensitivity in temperature  and polarization of this data combination.

Given that the main target of both future experiments is B-mode polarization from inflationary gravitational waves, we consider the combination of primordial tensors and PMFs. We first assume a fiducial model with r=0 and no PMFs, obtaining  $B_{1 \,\mathrm{Mpc}}<1 $ nG and $r<0.0004$ at 95\% CL. The result shows no signs of strong degeneracies, as shown by the two dimensional plot in the second panel of Fig.\ref{plotFuturo}. The limits on both r and the PMFs are compatible with the separate cases of PMFs and primordial tensor modes taken by themselves, with the limit on the PMFs amplitude dominated by the Simons Observatory data and the limit on the tensor to scalar ratio by the LiteBIRD data.
We then analyse the case with non-zero primordial tensor contribution, choosing a value of $r=0.004$. This value is compatible with theoretical predictions of $R^2$ model of inflation \cite{Starobinsky:1980te}, that is at the center of the region allowed by current data \cite{Akrami:2018odb} and can be detected with an high significance by the LiteBIRD experiment.
For this case we obtain $B_{1 \,\mathrm{Mpc}}<0.99 $ nG at 95\% CL and $r=0.0039\pm 0.0010$ always at 95 \% CL, which shows minimal correlation between the two parameters also in this case. The situation is different when considering the scale invariant PMFs configuration, in this case we obtain  $B_{1 \,\mathrm{Mpc}}<0.84 $ nG at 95\% CL and $r=0.0036\pm 0.0010$ always at 95 \% CL, which shows hints of a degeneracy brought by the spectral shape of magnetically induced B-mode for the scale invariant configuration as shown in the third panel of Fig.\ref{plotFuturo} also pointed out in \cite{Renzi:2018dbq}. 

We then explore the capabilities of detection of PMFs with future observations. By considering 1 nG and -2.9 as fiducial values for the fields amplitude and spectral index we have that the combination LiteBIRD+SO data combination is able to detect such PMFs at the level of $B_{1 \,\mathrm{Mpc}}=1.0\pm0.1$ 95\% CL.
In Table \ref{SOT} we summarize the main constraints for the combination LiteBIRD+SO.

\begin{table}[tbp]
\centering
\begin{tabular}{|c|c|c|c|c|}
\hline
Frequency GHz & FHWM [arcmin] & T Sensibility[$\mu K$ arcmin] & P Sensibility[$\mu K$ arcmin] \\
\hline
$93$  & 2.2 &5.8 & 8.2 \\
$145$ & 1.4 & 6.3& 8.9\\
\hline
\end{tabular}
\caption{\label{Tableso} Simons Observatory instrumental characteristics [Simons]}
\end{table}
\begin{table}[tbp]
\centering
\begin{tabular}{|l|c|c|c|c|c|c|}
\hline
Model& Baseline & $n_{B}=-2.9$ & $n_{B}=2$ & r=0 & r=0.004 & SI+r=0.004 \\
\hline
$B_{1 \,\mathrm{Mpc}}$ [nG] & < 1.0 & < 0.055& < $5\times 10^{-4}$& < 1.0& <0.99& <0.84 \\
$r$ & - & - & - &<0.0004 &$r=0.0039\pm 0.001$ & $r=0.0036\pm 0.001$ \\
\hline
\end{tabular}
\caption{\label{SOT} Constraints on the PMFs amplitude with the combination LiteBIRD+SO. All limits provided are at 95\% CL.}
\end{table}

\begin{table}[tbp]
\centering
\begin{tabular}{|c|c|c|c|c|}
\hline
Frequency GHz & FHWM [arcmin] & T Sensibility[$\mu K$ arcmin] & P Sensibility[$\mu K$ arcmin] \\
\hline
$85$  & 4.9 &1.77 & 2.5 \\
$95$ & 4.4 & 1.41&2.0\\
$145$ & 2.9 &1.84 &2.6\\
$155$ & 2.7 &1.91 &2.7\\
$215$ & 1.7  & 4.74&6.7\\
\hline
\end{tabular}
\caption{\label{Table3} Stage IV instrumental characteristics \cite{Calabrese:2016eii}}
\end{table}

\subsubsection{Combination with CMB Stage-IV}
We investigate some of the main forecasts also for the combination of LiteBIRD and S4 to get an overview of possible scenarios in a farther future. For the instrumental characteristics for a stage IV experiment we use \cite{Calabrese:2016eii} as reported in Table \ref{Table3}, once combined in inverse noise weighting we obtain an overall few microK sensitivity and 1-2' FWHM as adopted in S4 science book  \cite{Abazajian:2019eic,Abazajian:2016yjj}. We again assume a wide survey with the 40\% of the sky. 
For the baseline case of PMFs with the marginalization on the spectral index we obtain $B_{1 \,\mathrm{Mpc}}<0.63 $ nG with $n_{\mathrm B}<-0.63$ showing an improvement with respect to Simons Observatory thanks to the improved sensitivity. For the conservative delensing options considered we obtain again similar results to the unlensed case $B_{1 \,\mathrm{Mpc}}<0.61 $ nG for 20\% delensing, $B_{1 \,\mathrm{Mpc}}<0.59 $ nG for 40\% with only a minimal improvement for this case. 
For the specific PMFs configurations we have that the almost scale invariant is constrained to $B_{1 \,\mathrm{Mpc}}<0.49 $ nG without delensing, $B_{1 \,\mathrm{Mpc}}<0.45 $ nG for 40\% delensing and $B_{1 \,\mathrm{Mpc}}<0.40 $ nG for the extreme  80\% delensing case with again an improvement due to the delensing for the infrared indices configurations.
We tested again also the minimal spectral index for causal generated PMFs configurations which is limited to $B_{1 \,\mathrm{Mpc}}<0.2 $ pG showing a significant improvement with respect to the combination with Simons Observatory. 
Although the combination of LiteBIRD and Simons observatory constitutes  an improvement with respect to current results from Planck and ground based experiments, the improved sensitivity of CMB Stage IV will lead to even tighter bounds. 

\section{Conclusions}
The cosmological magnetic fields observed on large scales might have their roots in primordial seeds generated in the early Universe.The generation mechanisms are several and range from a first order phase transition to the breaking of conformal invariance during inflation opening a new observational window on the physics of the early Universe. PMFs being a fully relativistic component with anisotropic pressure diffuse on cosmological scales have an impact on the entire history of the Universe with several observational probes with an always increasing diversity thanks to the progress of cosmological observations. 

In this paper we have focused on one of the best known effects of PMFs: 
the gravitational effect on CMB anisotropies. We have upgraded our previous 
treatments using a completely new fitting technique for the exact solutions 
of the source terms of magnetically-induced perturbations reaching an unprecedented 
accuracy that improves theoretical predictions for the angular power spectra for 
current and future CMB experiments. Thanks to the increased numerical stability 
due to both improved fits and improved basic camb version we could also increase 
the maximum multipole of the tensor modes to 2000 taking more advantage of the high multipole part of the passive tensor mode for positive spectral indices.

We have updated the constraints on PMFs to Planck 2018 data. Using the Planck 2018 baseline alone we obtain $B_{1 \,\mathrm{Mpc}}<3.4$ nG with the addition BICEP/KECK data we have $B_{1 \,\mathrm{Mpc}}<3.5$ nG and $B_{1 \,\mathrm{Mpc}}<2.9$ nG when adding also the South Pole Telescope, all at 95\% C.L.. The improvement with respect to previous results with Planck 2015 \cite{Ade:2015cva} is mainly due to the combination of improved data on polarization by Planck 2018 and the improvement in our methodology. The results for specific configurations of the PMFs are the following: for $n_{\mathrm B}=2$ is $B_{1 \,\mathrm{Mpc}}<3$ pG at 95\% C.L. for the combination of Planck, BICEP/Keck and SPT; for the almost scale invariant case, that with its infrared shape represents the maximum contribution on large angular scales in polarization, we found a looser bound  $B_{1 \,\mathrm{Mpc}}=1.5^{+0.9}_{-1.4}$ nG at 95\% C.L. 
The addition of primordial inflationary tensor mode does not impact the results, and both	 the constraints on the PMFs amplitude and tensor to scalar ratio are unchanged, showing a lack of correlation of the two signals for the  precision of current data. A non-vanishing correlation is present for the almost scale invariant case due to the similarity of contributions on large angular scales but does not affect our results significantly.

We have then performed the forecasts for future experiments. In particular, we have considered the latest 
LiteBIRD instrumental configuration \cite{Hazumi:2019lys} showing how forecasted constraints are at the 
Planck level, i.e.  $B_{1 \,\mathrm{Mpc}}<3.8$ nG at 95\% C.L., when considering only temperature and 
E-mode polarization and improve significantly when adding B-mode polarization, i.e. $B_{1 \,\mathrm{Mpc}}<2.7$ nG at 95\% C.L.
The combination of LiteBIRD and ground based experiments as Simons Observatory and CMB Stage IV brings the constraints for generic 
PMFs configuration {\it below} nanoGauss level. The effect of delensing is mitigated when the index of PMFs is allow to vary, but 
is important for specific spectral indices. The constraints on the amplitude with $n_{\mathrm B}=2$ will be improved by more than 
one order of magnitude because the vector contribution will be better constrained by the future ground based experiments. The almost 
scale invariant configuration is very interesting:  the major improvement comes for LiteBIRD sensitivity at large and intermediate 
angular scales, and in addition the constraints will improve for large delensing factors. This case is also interesting since it 
shows a slight correlation between the PMFs contribution and the primary inflationary B-modes.

We have shown that for a configuration of the PMFs with the spectral index free to vary, the gravitational 
impact on CMB anisotropies constrains the amplitude at the level of few nanoGauss with current data from 
Planck 2018, BICEP/Keck and SPT. Future data from LiteBIRD, together with its combination with Simons Observatory 
or Stage IV, will be able to break this level and significantly improve the constraints by almost an order of magnitude. 
The situation further improves when we consider specific PMFs configurations. The strong impact on small angular 
scales of causal fields, $n_{\mathrm B}\ge 2$, improves the constraints by orders of magnitude, reaching the sub-picoGauss 
level for the future ground based experiments. This results will squeeze the allowed observational window given by the 
upper limits from the CMB and the lower limits from gamma-ray observations.
The almost scale invariant configuration is interesting for two reasons. 
On one side LiteBIRD will provide much tighter constraints than current data, 
and on the other side this type of PMF contribution shows a correlation with 
the primordial gravitational wave signal. Therefore, it represents a possible 
source of confusion to be taken in account for future high precision data.

Our analysis considered semi-idealistic conditions with a perfect cleaning of 
the foregrounds but ongoing studies for future experiments (e.g.\cite{Stompor:2016hhw}) 
show that there might still be some residual contamination on large angular scales even after cleaning. 
Previous studies on small angular scales have already shown how the analogous residuals degrade the constraints 
on PMFs  amplitude \cite{Paoletti:2012bb,Ade:2015cva}, the update of these studies on large angular scale 
polarization in the context future experiments is in progress. It will be interesting to study the
status and  perspectives for another key signature from PMFs on CMB
polarization as the one from the post-recombination heating.

\acknowledgments
DP and FF acknowledge financial support by ASI Grant 2016-24-H.0.
This research used computational resources provided by INAF OAS Bologna and by CINECA under the agreement with INFN.

\end{document}